# Growth strategy of microbes on mixed carbon sources

Xin Wang[1,2,†], Kang Xia[1,3,†], Xiaojing Yang[1], Chao Tang[1,*]

[1]Center for Quantitative Biology, School of Physics and Peking-Tsinghua Center for Life Sciences, Peking University, Beijing 100871, China.

[2]Channing Division of Network Medicine, Brigham and Women's Hospital and Harvard Medical School, Boston, Massachusetts 02115, USA.

[3]College of Life Sciences, Wuhan University, Wuhan 430072, China.

[†]These authors contributed equally to this work.

[*]Correspondence and requests for materials should be addressed to C.T. (email: tangc@pku.edu.cn).

**Abstract**

A classic problem in microbiology is that bacteria display two types of growth behavior when cultured on a mixture of two carbon sources: the two sources are sequentially consumed one after another (diauxie) or they are simultaneously consumed (co-utilization). The search for the molecular mechanism of diauxie led to the discovery of the *lac* operon. However, questions remain as why microbes would bother to have different strategies of taking up nutrients. Here we show that diauxie versus co-utilization can be understood from the topological features of the metabolic network. A model of optimal allocation of protein resources quantitatively explains why and how the cell makes the choice. In case of co-utilization, the model predicts the percentage of each carbon source in supplying the amino acid pools, which is quantitatively verified by experiments. Our work solves a long-standing puzzle and provides a quantitative framework for the carbon source utilization of microbes.

During the course of evolution, biological systems have acquired a myriad of strategies to adapt to their environments. A great challenge is to understand the rationale of these strategies on quantitative bases. It has long been discovered that the production of digestive enzymes in a microorganism depends on (adapts to) the composition of the medium[1]. More precisely, in the 1940s Jacques Monod observed two distinct strategies in bacteria (*E. coli* and *B. subtilis*) to take up nutrients. He cultured these bacteria on a mixture of two carbon sources, and found that for certain mixtures the bacteria consume both nutrients simultaneously while for other mixtures they consume the two nutrients one after another[2, 3]. The latter case resulted in a growth curve consisted of two consecutive exponentials, for which he termed this phenomenon diauxie. Subsequent studies revealed that the two types of growth behavior, diauxic- and co-utilization of carbon sources are common in microorganisms[4-8]. The regulatory mechanism responsible for diauxie, that is the molecular mechanism for the microbes to express only the enzymes for the preferred carbon source even when multiple sources are present, is commonly ascribed to catabolite repression[5, 9-13]. In bacteria it is exemplified by the *lac* operon and the cAMP-CRP system[14-17]. In yeast, the molecular implementation of catabolite repression differs, but the logic and the outcome are similar[5].

Why have microbes evolved to possess the two strategies and what are the determining factors for them to choose one versus the other? For unicellular organisms, long term survival and growth at the population level are paramount. Cells allocate their cellular resources to achieve optimal growth[18-27]. In particular, it has been demonstrated that the principle of optimal protein resource allocation can quantitatively explain a large body of experimental data[20, 21] and that the most efficient enzyme allocation in metabolic networks corresponds to elementary flux mode[25, 26, 28]. In this paper, we extend these approaches to address the question of multiple carbon sources and show that the two growth strategies can be understood from optimal enzyme allocation further constrained by the topological features of the metabolic network.

## Results

### Categorization of Carbon Sources

Carbon sources taken by the cell serve as substrates of the metabolic network, in which they are broken down to supply pools of amino acids and other components that make up a cell. Amino acids take up a majority of carbon supply (about 55%)[29-31]. As shown in Fig. 1, different carbon sources enter the metabolic network at different points[31]. Denote those sources entering the upper part of the glycolysis Group A and those joining at other points of the metabolic network Group B (Fig. 1). Studies have shown that when mixing a carbon source of Group A with that of Group B, the bacteria tend to co-utilize both sources and the growth rate is higher than that with each individual source[6, 7, 32]. When mixing two sources both from Group A, the bacteria usually use a preferred source (of higher growth rate) first[4, 6, 11, 13, 33, 34].

**Precursor Pools of Biomass Components**

Based on the topology of metabolic network (Fig. 1), we classify the precursors of biomass components (amino acids and others) into seven precursor pools. Specifically, each pool is named depending on its entrance point on the metabolic network (see Supplementary Note 1.3 for details): a1 (entering from G6P/F6P), a2 (entering from GA3P), a3 (entering from 3PG), a4 (entering from PEP), b (entering from pyruvate/Acetyle-CoA), c (entering from α-Ketoglutarate) and d (entering from oxaloacetate). The Pools a1-a4 are collective called as Pool a.

**Coarse-graining the metabolic network**

Note that the carbon sources from Group A converge to the node (G6P/F6P) before entering various pools, while the carbon sources from Group B can take other routes (Fig. 1). In fact, the metabolic network shown in Fig. 1 can be coarse-grained (see Methods) into a model shown in Fig. 2a, in which nodes $A1$ and $A2$ represent carbon sources from Group A, node $B$ a source from Group B, and Pools 1 and 2 are some combinations of the four pools in Fig. 1. In other words, Fig. 2a is topologically equivalent to Fig. 1 as far as the carbon flux is concerned. Each coarse-grained arrow (which can contain several metabolic steps) carries a carbon flux $J$ and is characterized by two quantities: the total enzyme cost $\Phi$ dedicated to carry the flux and a parameter $\kappa$ so that $J = \Phi \cdot \kappa$.

**Origin of Diauxie for Carbon Sources in Group A**

Let us first consider the case in which both carbon sources are from Group A. We proceed to solve the simple model of Fig. 2a with two sources $A1$ and $A2$ ($[B]=0$), using the optimal protein allocation hypothesis[18, 20, 21, 25], which maximizes the enzyme utilization efficiency.

In Fig. 2a, all enzymes that carry and digest nutrient $Ai$ ($i = 1, 2$) into node $M$ are simplified to a single effective enzyme $E_{Ai}$ of cost $\Phi_{Ai}$ (see Supplementary Note 1.2 for details). The carbon flux to the precursor pools from source $Ai$ is proportional to $\Phi_{Ai}$. We take the Michaelis-Menten form (see Supplementary Note 1.2 for details):

$J_{Ai} = \Phi_{Ai} \cdot \kappa_{Ai}$, where $\kappa_{Ai} = k_{Ai} \cdot \frac{[Ai]}{[Ai] + K_{Ai}}$ (denoted as the substrate quality). $[Ai]$ is the concentration of $Ai$. For the subsystem consist of $A1$, $A2$ and node $M$, $J_{tot} = \Phi_{A1} \cdot \kappa_{A1} + \Phi_{A2} \cdot \kappa_{A2}$ and $\Phi_{tot} = \Phi_{A1} + \Phi_{A2}$. We define the efficiency of a pathway by the flux delivered per total enzyme cost[25, 26]:

$$\varepsilon \equiv \frac{J_{tot}}{\Phi_{tot}}. \qquad (1)$$

The efficiency to deliver carbon flux from the two sources $A1$ and $A2$ to node $M$ is then $\varepsilon = \frac{\Phi_{A1} \cdot \kappa_{A1} + \Phi_{A2} \cdot \kappa_{A2}}{\Phi_{A1} + \Phi_{A2}}$. If $\kappa_{A1} > \kappa_{A2}$, that is, if the substrate quality of $A1$ is better than that of $A2$, then $\varepsilon = \kappa_{A1} - \frac{\Phi_{A2} \cdot (\kappa_{A1} - \kappa_{A2})}{\Phi_{A1} + \Phi_{A2}} \leq \kappa_{A1}$. It is easy to see that the optimal solution (maximum efficiency) is $\Phi_{A2} = 0$. This means that the cell expresses only the enzyme for $A1$ and thus utilizes only $A1$. Conversely, if $\kappa_{A1} < \kappa_{A2}$, $\Phi_{A1} = 0$ is optimal and the cell would utilize $A2$ only. In either case, optimal growth would imply that cells only consume the preferable carbon source, which corresponds to the case of diauxie[2, 3, 6, 9, 11].

In the above coarse-grained model, the enzyme efficiency of the carbon source $Ai$ is lump summed in a single effective parameter $\kappa_{Ai}$. In practice, there are intermediate nodes and enzymes along the pathway as depicted in Fig. 2b, and more elaborate calculations taking into account the intermediate steps are needed to evaluate the pathway efficiency. Note that Fig. 2b is rather generic in representing a part of the metabolic pathway under consideration. $X$ and $Y$ can represent carbon sources coming from Group A and/or Group B, $M$ represents the convergent node of the two sources under consideration and Pool z represents the precursor pool under consideration. We now proceed to calculate and compare the efficiencies of the two branches: $X \rightarrow M$ and $Y \rightarrow M$. Using the branch $X \rightarrow M$ as the example, denote $E_X^j$ the enzymes (of protein cost $\Phi_X^j$) catalyzing the intermediate nodes $m_X^j$ ($j = 1 \sim N_X$) (Fig. 2b). Define $\Phi_X^{\text{b}} \equiv \Phi_X + \sum_{j=1}^{N_X} \Phi_X^j$, which is the total protein cost for enzymes dedicated to the branch. The pathway efficiency for the branch $X \to M$ is then $\varepsilon_{X \to M} = J_{X \to M} / \Phi_X^{\text{b}}$, where $J_{X \to M}$ is the carbon flux from $X$ to $M$. Assuming that the flux is conserved in each step along the branch, $J_{X \to M} = \Phi_X \cdot \kappa_X = \Phi_X^j \cdot \kappa_X^j$ ($j = 1 \sim N_X$), where $\kappa_X^j = k_X^j \cdot \frac{[m_X^j]}{[m_X^j] + K_X^j} \approx k_X^j$ is the substrate quality of $m_X^j$ (Supplementary Note 1.2) and the last approximation is valid with $[m_X^j] \geq K_X^j$ which is generally true in bacteria[35, 36] and which also maximizes the flux with a given enzyme cost (Supplementary Note 1.5). It is then easy to see that

$$\varepsilon_{X \to M} = \frac{1}{1/\kappa_X + \sum_{j=1}^{N_X} 1/\kappa_X^j} \approx \frac{1}{1/\kappa_X + \sum_{j=1}^{N_X} 1/k_X^j}. \qquad (2)$$

For $\varepsilon_{X \to M} > \varepsilon_{Y \to M}$, the optimal solution is $\Phi_Y = \Phi_Y^j = 0$; for $\varepsilon_{X \to M} < \varepsilon_{Y \to M}$, the optimal solution is $\Phi_X = \Phi_X^j = 0$ (see Supplementary Note 2.2 for details). Only the nutrient with higher branch efficiency is utilized to supply the convergent node $M$ and thus Pool z (Fig. 2b).

When $X$ and $Y$ both come from Group A, the convergent node $M$ resides upstream to all precursor pools (Fig. 1 and Supplementary Fig. 1c). Only the nutrient with higher efficiency is being utilized to supply all precursor pools. When the preferred nutrient is exhausted, the cell switch to the other less favorable nutrient. The actual switching point could depend on the concentrations of both nutrients. Note that the branch efficiency $\varepsilon_{Ai \to M}$ (Eq. (2)) depends on the concentration of the nutrient $[Ai]$ through the substrate quality $\kappa_{Ai} = k_{Ai} \cdot \frac{[Ai]}{[Ai] + K_{Ai}}$. Thus, only with saturating concentrations, one can have an absolute ranking of the nutrient quality. For concentration $[Ai] < K_{Ai}$, $\varepsilon_{Ai \to M}([Ai])$ drops fast with $[Ai]$. Theoretically, when the concentration of the originally preferred nutrient (A1) drops to a point $[A1]_{\text{T}}$ such that $\varepsilon_{A1 \to M}([A1]_{\text{T}}) = \varepsilon_{A2 \to M}([A2])$, the other nutrient (*A2*) becomes preferrable, and the cell may switch to *A2* at this point. This gives $[A1]_T = \dfrac{\delta \cdot [A2]}{\Delta + [A2]}$, where $\delta$ and $\Delta$ are constant (see Supplementary Note 2.3 for details). For $[A2] \ll \Delta$, the turning point is reduced to $[A1]_T = \dfrac{\delta}{\Delta} \cdot [A2]$, a form of ratio sensing. Indeed, ratio sensing was recently observed in the budding yeast *Saccharomyces cerevisiae* cultured in glucose-galactose mixed medium[33], and the experimental results agree well with the turning point equation derived above (see Supplementary Fig. 2 and Supplementary Note 2.3 for details).

**Co-Utilization of Carbon Sources**

The diauxic growth is due to the topology of the metabolic network, in which Group A sources converge to a common node (G6P/F6P) before diverting to various precursor pools (Figs. 1, 2a and Supplementary Fig. 1c). The situation is different if the two mixed carbon sources are from Groups A and B, respectively (denoted as A+B). (Some combinations of two Group B sources also fall into this category and can be analyzed similarly; see Supplementary Fig. 3d.) Group B sources can directly supply some precursor pools without going through the common node (G6P/F6P) (Fig. 1). The topologies of the metabolic network in the cases of A+B are exemplified in Supplementary Fig. 3. All A+B cases can be mapped to a common coarse-grained model depicted in Supplementary Fig. 1d (which is also Fig. 2a with only one of the A sources present), although the actual position of nodes $M$ and $N$ in the metabolic network, and the contents of Pools 1 and 2 may depend on specific cases. As obvious from Fig. 2a, source *A* or *B* alone could in principle supply all precursor pools. However, because of the location of the precursor pools relative to the sources, it may be more economical for one pool to draw carbon flux from one source and the other from the other source.

To determine which of the two carbon sources should supply which pool(s), we apply branch efficiency analysis (see Supplementary Note 3 for more details). For Pool 1, we compare the efficiency of *A* and *B* in supplying flux to node $M$; while for Pool 2 to node $N$. The criteria are simply:

$$\text{Pool 1 is supplied by } \begin{cases} A, \text{ if } \varepsilon_{A \to M} > \varepsilon_{B \to M} \\ B, \text{ if } \varepsilon_{A \to M} < \varepsilon_{B \to M} \end{cases}. \quad (3)$$

$$\text{Pool 2 is supplied by } \begin{cases} A, \text{ if } \varepsilon_{A \to N} > \varepsilon_{B \to N} \\ B, \text{ if } \varepsilon_{A \to N} < \varepsilon_{B \to N} \end{cases}. \quad (4)$$

It is easy to see from inequalities (3) and (4) that if the following condition is met

$$1/\kappa_B - 1/\kappa'_M < 1/\kappa_A < 1/\kappa_B + 1/\kappa'_N, \qquad (5)$$

then *A* supplies Pool 1 and *B* supplies Pool 2 -- the two carbon sources are simultaneously consumed. In reality, there are multiple intermediate nodes between the $M$ - $N$ interconversion (Figs. 1, 2a and Supplementary Fig. 1d). Similar to Eq. (2), $1/\kappa'_M$ and $1/\kappa'_N$ here actually represent summations of all intermediate terms between $M$ and $N$ in the metabolic network.

**Pools Suppliers in the Mixed Carbon Sources**

In order to apply the above analysis to the real case, we collected the available data for metabolic enzymes of *E. coli* from the literature (Supplementary Table 1). We calculated the branch efficiencies of different carbon sources to the metabolites F6P, GA3P, 3PG, PEP, pyruvate and oxaloacetate (see Supplementary Note 4.1 for details), which correspond to the nodes $M$ or $N$ in the simplified network of Fig. 2a for Pools a1-d (Fig. 1 and Supplementary Fig. 3). The results are shown in Supplementary Table 2. Then using the criteria Eqs. (3) and (4), one could evaluate the carbon source supplier(s) of all pools (Supplementary Table 3). Note that Pool c is supplied by both the suppliers of Pools b and d owing to the effect of converged flux (see Supplementary Notes 4.1-4.2 for details).

However, in practice, the suppliers of Pools b-d can be different from the above evaluation due to energy production in the TCA cycle. Specifically, when oxaloacetate flow through a TCA cycle back to itself, it generates fixed amount of energy[31], with half of the carbon atoms replaced by those coming from pyruvate (see Supplementary Figs. 4b-c and Supplementary Notes 4.3-4.4 for details). By collecting relevant energy production data from literatures[37], we quantitatively analyzed the influence of this effect, and obtained the optimal carbon source supplier(s) of each precursor pool for *E. coli* under aerobic growth in various combinations of source mixture (see Supplementary Notes 4.3-4.5 for details). The results are shown in Supplementary Table 4.

**Comparison with experiments**

To test these predictions (Supplementary Table 4), we use $^{13}C$ isotope labeling methods to trace the carbon source(s) of each precursor pool (see Methods for details). Specifically, we cultured *E. coli* to the steady state in a mixture of two carbon sources with one source being labeled with $^{13}$C. We then measured the $^{13}$C labelling percentage of amino acids and obtained the percentage of each carbon source in supplying the synthesis of each amino acid. To ensure reliability, we obtained and compared our experimental results using two types of fragment in mass spec raw data: M-57 and M-85 (Supplementary Fig. 5) (see Methods for details).

We first examined the A+B cases (Group A source: glucose, lactose, fructose, glycerol; Group B source: pyruvate, succinate, fumarate, malate). Overall, the experimental results showed excellent agreement with our predictions (Fig. 3). A number of features are worth noting. Just as the model predicted (Supplementary Table 4), two patterns of the carbon source partition (Figs. 3a-b) were observed depending on which A source was used. Glucose and lactose are both highly preferable carbon sources for *E. coli,* both supporting large growth rates (Supplementary Table 5); their supply patterns look almost the same (Fig. 3a). Fructose, glycerol, maltose and galactose are less preferable Group A sources with lower growth rates (Supplementary Table 5), and they showed very similar supply patterns when mixed with the same Group B source (Fig. 3b, Supplementary Fig. 6a). Group B sources succinate, fumarate and malate showed similar supply patterns when mixed with a same Group A source (Figs. 3a-b, 4a & Supplementary Fig. 6a). There was a noticeable systematic discrepancy between the experimental results and the model predictions for Pool a. This may be due to the fact that microbes reserve a portion of gluconeogenesis enzymes preparing for potential changing environment (see also Supplementary Notes 4.8-4.9).

Next, as our model can calculate carbon source utilization and partition in any combinations of sources, we performed experiments for B+B cases (pyruvate mixed with succinate, fumarate or malate; succinate mixed with malate). In agreement with the model prediction, these B+B cases showed co-utilization and the measured carbon supply percentages quantitatively agree with model predictions (Fig. 4).

## Discussion

The diauxie versus co-utilization puzzle can be understood from the topology of the metabolic network. This can be illustrated with the coarse-grained model shown in Fig. 2a (see also Supplementary Figs. 1b-d). The sources of Group A go through a common node before delivered to various precursor pools, and the most efficient source wins[22]. It has been observed that there is a hierarchy among Group A sources ranked according to the single-source growth rate, and when two or more sources are present the bacteria usually use the one with highest growth rate[6, 34]. This is a natural consequence of our theory. A higher growth rate commonly implies higher enzyme utilization efficiency and thus a higher priority to be utilized. Other than the *lac* operon, questions remain as how this priority is implemented molecularly. It has been known that in many cases the catabolite repression is not complete and that this may depend on whether the carbon sources belong to the type of Phosphotransferase System (PTS)[10, 13], highlighting the potential constraints, trade-offs and/or costs of implementing a prefect optimal solution. We have mixed glucose (a PTS sugar) with both PTS sugar fructose and non-PTS sugars maltose and glycerol, which all belong to the A+A cases. We found that while glucose showed almost perfect inhibition to the two non-PTS sugars (all precursor pools were about 100% supplied by glucose), its inhibition to the other PTS sugar fructose was not complete (glucose supplied ~83% precursor pools) (Supplementary Fig. 6b).

When Group B source is present along with Group A source, it can take a shortcut to reach some of the precursor pools (Fig. 2a, Supplementary Fig. 1d) and can be more efficient to supply these pools. Some combinations of two Group B sources also fall into this category and thus can be co-utilized. In these cases, our experimental results quantitatively agree with our model predictions. As can be seen from Figs. 3 and 4, despite the various possible combinations of carbon sources, the partition of the sources among the pools fall into a few patterns. This is due to the fact that these partitions are largely determined by the topology of the metabolic network and thus are quantized. This property also relaxes the requirements of accurate enzyme parameters in determining the pool suppliers. To test the robustness of the model predictions with respect to the errors/uncertainties of the parameters extracted/estimated from the literature, we carried out a detailed analysis (Supplementary Note 4.9 and Supplementary Tables 6 and 7). The analysis showed that for any mixture of two carbon sources and for arbitrary choice of parameters, only a very few (no more than 4) partition patterns of the sources are qualitatively similar to the experimental result. Model predictions using the nominal parameter values from the literature quantitatively and consistently agree with all the experimental patterns for all the combinations of carbon sources we tested. Conversely, in order to produce a pattern that is qualitatively similar but quantitatively different from the experimental one, a very large deviation from at least one nominal value is necessary.

The present work deals with relatively stable growth conditions and the simple exponential growth behavior. In this case, there is a body of experimental evidence for optimal protein allocation[18-24, 27, 38]. Furthermore, our model relies only on the assumption that microbes optimize enzyme utilization efficiency, so it may also be applicable to suboptimal growth cases[27, 38]. The environment the microbes face can be highly variable and uncertain. Their long-term fitness of the population may not simply be determined only by the growth rate of individual cells in the exponential phase, but a result of trade-offs that best adapt to the changing environment. Strategies such as bet hedging, memory of the past and anticipation of the future are found to exist in microorganisms[39-48].

Finally, from theoretical aspects, our analysis framework is broadly applicable to more complex regulations in metabolic networks such as reversible reactions, allosteric enzymes, metabolites inhibitions, etc. (see Supplementary Notes 5-7 for details). However, there are cases, such as bi-substrates transporters or enzymes (e.g. glucose transporters in *E. coli* can co-transport mannose[49]), for which specific care needs to be taken (see Supplementary Notes 6-7 for details). In practice, the nutrient uptake strategy or eating habit of a microbe is shaped by its environmental history. While the phenomena of diauxie versus co-utilization are widely spread in microbes, there are bound to be variations and exceptions. For example, certain microbes may have different hierarchies of preferable carbon sources[4]. It is still a great challenge to understand in quantitative frameworks how cells and population behave and evolve in different and changing environments.

## Methods

**Coarse graining methods**

Coarse graining of the metabolic network is done in such a way as to preserve the network topology but grouping metabolites, enzymes and pathways into single representative nodes and corresponding effective enzymes. In particular, a linear pathway is lump summed into two nodes (start and end) connecting with a single effective enzyme.

**Strain.**

The strain used in this study is *E. coli* K-12 strain NCM3722.

**Growth medium:** Most of the cultures were based on the M9 minimal medium (42mM $Na_2HPO_4$, 22mM $KH_2PO_4$, 8.5mM NaCl, 18.7mM $NH_4Cl$, 2mM $MgSO_4$, 0.1mM $CaCl_2$), and supplemented with one or two types of carbon sources. For the carbon sources in each medium, the following concentrations were applied: 0.4% (w/v) glucose, 0.4% (w/v) lactose, 0.4% (v/v) glycerol, 20 mM fructose, 20 mM fructose, 20 mM maltose, 20 mM pyruvate, 15 mM succinate, 20 mM fumarate, and 20 mM malate.

**Batch culture growth:** The batch cultures were performed either in the 37°C incubator shaker shaking at 250 rpm or in the microplate reader, which holds the temperature at 37°C and shakes at 900 rpm. The culture volume was 200 μL in 96-well plates, 1 mL in 5 mL round-bottom tubes or 50 mL in 100 mL flasks. Every batch culture was performed as described below. Single colony from the LB agar plate was first inoculated into 50 mL LB medium and cultured overnight. Then, 0.5 mL overnight culture was inoculated into 50 mL LB medium and cultured for 2 hours. Cells were centrifuged at 4,000 rpm for 2 minutes, and the cell pellets were diluted to $OD_{600} = 0.001$ in the culture medium (M9 medium supplemented with various carbon sources). Then, the medium was cultured in microplate reader (measuring growth rate) or incubator shaker (isotope labeling).

**Growth rate measurement:** Each well of the 96-well plate was covered with a 200 μL culture medium ($OD_{600} = 0.001$). Cells were cultured at 37°C in the microplate reader shaking at 720 rpm with a 2 mm diameter. The microplate reader measured the $OD_{600}$ of each well at an interval of 5 minutes for 20 hours. Growth rate $\lambda$ was measured as the multiplicative inverse of doubling time:

$$\lambda = \frac{d \log_2 OD_{600}}{dt}. \tag{6}$$

The interval of 2 hours with the maximum Pearson correlation coefficient was defined to be in the exponential phase and the slope of this interval was the growth rate. For some cultures, the $OD_{600}$ remains constant during the first 20 hours. Thus the record time was extended to 72 hours for these cultures.

**Isotope labeling:** The following $^{13}C$ carbon sources were applied in the isotope labeling experiments: glucose (Product code: CLM-1396; Cambridge Isotope Laboratories, Inc.), fructose (Product number: 587621; Sigma-Aldrich) and glycerol (Product number: 489476; Sigma-Aldrich) (Group A); pyruvate (Product code: CLM-2440; Cambridge Isotope Laboratories, Inc.) and succinate (Product number: 491985; Sigma-Aldrich) (Group B). In an isotope labeling experiment, there are two types of carbon sources: one is uniformly labeled with $^{13}C$, while the other one is not labeled. In every experiment, 1 mL $^{13}C$-labeled culture medium ($OD_{600} = 0.001$) was inoculated into the 5 mL round-bottom tube and cultured in the incubator shaker until the $OD_{600} = 0.150$ to $0.250$ (7~8 generations). Three independent experiments (with numerous distinct cells) were carried out for each combination of mixed carbon sources.

**Extraction and derivatization of amino acids[50]:** Cells labeled by $^{13}C$ were harvested by centrifuging for 3 minutes at 12,000 rpm. The cell pellets were washed with 1 mL PBS and centrifuged for 2 minutes at 12,000 rpm twice, and then resuspended in 200 μL of 6

M HCl. The resuspended cells were transferred into sealed 1.5 mL tubes and hydrolysed for 20 hours at 105°C. The cell hydrolysate was dried at 65°C under the fume hood. The dry hydrolysate was resuspended with 40 μL *N,N*-Dimethylformamide and 20 μL *N*-tert-butyldimethylsilyl-*N*-methyltrifluoroacetamide and heated at 85°C for one hour so that the amino acids were derivatized (structure shown in Supplementary Fig. 7b). The solution of mixed derivatized amino acids was filtered with 13 mm syringe filter with 0.2 μm membrane.

**GC-MS setup:** GC-MS analysis was carried out using the Hybrid Quadrupole-Orbitrap GC-MS/MS System (Q Exactive GC, ThermoFisher). The injected sample volume was 1 μL at a carrier gas flow of 1.200 mL/min helium with a split ratio of 1:4.2. The oven temperature was initially set at 150 °C and maintained for 2 minutes, raised to 180 °C at 5 °C/min and immediately to 260 °C at 10 °C/min and maintained for 8 minutes, and then raised to 350 °C/min and maintained for 5 minutes. The ionization mode was set as electron impact ionization. The ion source temperature was set at 230 °C. The MS transfer line temperature was set at 250 °C. The scan range was 50.0 to 650.0 m/z with a resolution at 60,000. The MS was tuned to 414.0 m/z.

**GC-MS analysis of $^{13}$C labeled derivatized amino acids:** The derivatized amino acids was analyzed by GC-MS with the setup described above. Different kinds of derivatized amino acids in one sample were separated in the gas chromatography (GC) according to their different retention time. The amino acids were fragmented during ionization, forming different kinds of fragments (Supplementary Fig. 7b). Different kinds of fragment of the same derivatized amino acid and their relative abundance (Supplementary Figs. 7b-c) were analyzed by mass spectrometry (MS). The labeling percentage of a certain amino acid can be inferred from the labeling percentage of its fragments. Thermo Xcalibur4.0 was used to view and process the GC-MS data. According to the relative retention time in the chromatogram (Supplementary Fig. 7a) and the corresponding mass spectrum (Supplementary Fig. 7b), 13 kinds of amino acid were detected. The integrated mass spectrum over the full peak range of every derivatized amino acid was obtained to calculate the $^{13}$C labeling percentage[51]. For a typical derivatized amino acid, 5 types of fragment, M-15, M-57, M-85, M-159 and f302 (Supplementary Fig. 7b), were detected. M-57 denotes that this fragment weighs 57 daltons less than the corresponding derivatized amino acids, and the same goes for M-15, M-85 and M-159. f302 denotes a fragment of weights 302 daltons. For a certain fragment containing N carbon atoms from the underivatized amino acids (the natural form of concern), there were ($N+1$) kinds of mass isotopomer incorporating $0 \sim N$ $^{13}C$ respectively. $I_i$ denoted the intensity of mass isotopomer that had $i$ $^{13}C$ and $N-i$ $^{12}C$. The $^{13}C$ labeling percentage $\zeta$ of this fragment was calculated as follows:

$$\zeta = \frac{\sum_{i=1}^{N} i \cdot I_i}{N \cdot \sum_{i=1}^{N} I_i}. \qquad (7)$$

**Amino acids $^{13}C$ labeling data of different fragments:** There are 5 types of fragment (M-15, M-57, M-85, M-159 and f302) formed during the ionization of amino acids. With each type of raw data, we can calculate (with Eq. (7)) a set of $^{13}C$ labeling percentages for amino acids. According to the molecular structure, these results can be classified into 3 categories: using 1) M-15/M-57; 2) M-85/M-159; and 3) f302. Only in the first category, there is no carbon atom loss during the fragmentation of the amino acids of concern (Supplementary Fig. 7). This means that M-15/M-57 reflects the exact $^{13}C$ labeling percentages of amino acids. Yet, the signal intensity of M-15 is faint, thus M-57 is the best choice among the fragments. However, in practice, Leu M-57 and Ile M-57 fragments share the same mass as that of f302 and thus are not applicable for analysis. Consequently, we used Leu M-15 and Ile M-15 to calculate the $^{13}C$ labeling percentage of Leu and Ile, and M-57 data for other amino acids throughout our manuscript and Supplementary Information unless otherwise specified. In the second category, the carboxylic carbon atom in an amino acid was lost during the fragmentation (Supplementary Fig. 7). Thus M-85/M-159 data can reflect the $^{13}C$ labeling percentages for a majority yet not all carbon atoms in amino acids, and indeed M-85 and M57 data show a very good agreement with each other (Supplementary Fig. 5). In the third category, all carbon atoms in the side chain of an amino acid (a considerable proportion) were lost during fragmentation (Supplementary Fig. 7). As a result, we did not use f302 data for calculation in this study.

## Code availability

This paper does not involve computer code. Built-in functions of Origin (v9) were used for curve fitting in Supplementary Fig. 2.

## Data availability

The data that support the findings of this study are available from the corresponding author (C.T.) upon request. The source data underlying Figs. 3-4, Supplementary Figs. 5-6, Supplementary Note 4.7 and Supplementary Table 5 are provided with the paper.

## Acknowledgements

We thank Yiping Wang for providing us the strain NCM3722, and Terence Hwa, Yuan Yuan, Haoyuan Sun, Xiaohui Liu, Yanjun Li, Shaoqi Zhu and Yimiao Qu for helpful discussions. This work was supported by Chinese Ministry of Science and Technology (2015CB910300) and National Natural Science Foundation of China (91430217).

## Author Contributions

X.W. and C.T. conceived and designed the project, developed the model, designed the experiments and wrote the paper. X.W. carried out the theoretical analysis. K.X. carried out the experiments. X.W. and K.X. analyzed the experimental data. C.T. supervised the whole project. X.Y. co-supervised the experiments.

**Competing interests**: The authors declare no competing interests.

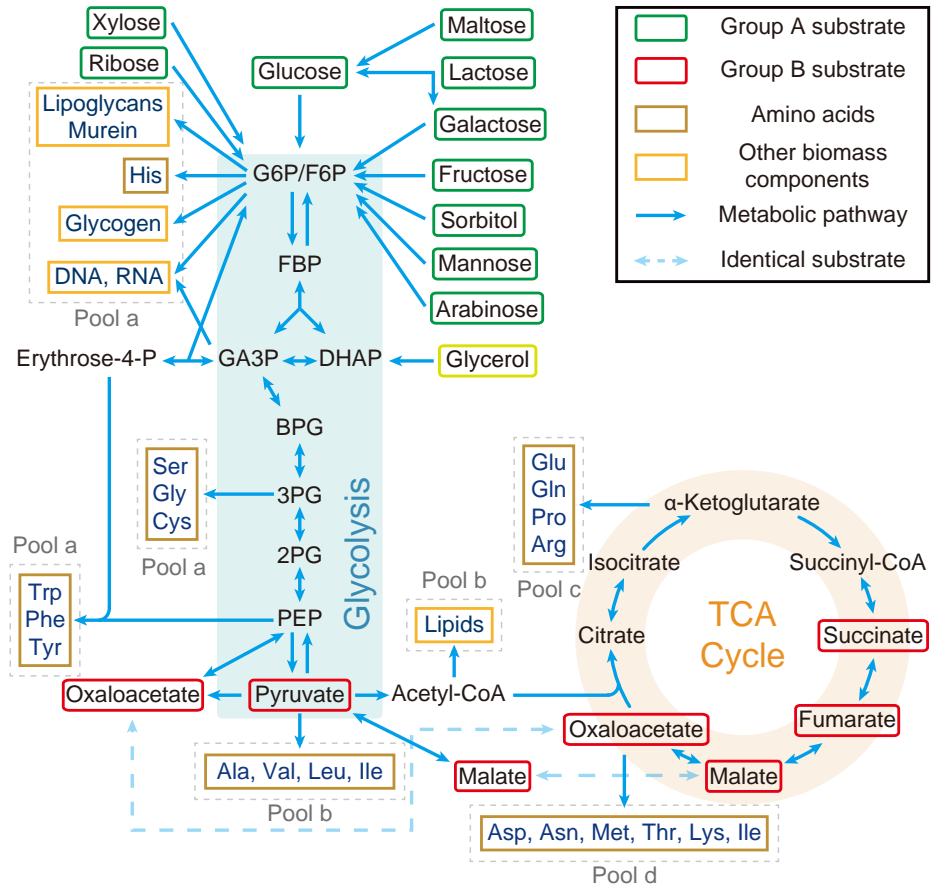

**Figure 1 | Metabolic network of carbon source utilization.** Group A substrates (in green squares) can be simultaneously utilized with Group B substrates (red squares), whereas substrates paired from Group A usually display diauxie. Only the major pathways are shown. The precursors of biomass components (amino acids marked with light brown squares and other components marked with orange squares) are classified into Pools a-d (marked with grey dashed line squares). The enzyme for the interconversion between Glucose 6-phosphate (G6P) and fructose 6-phosphate (F6P) is very efficient (Supplementary Table 1), so we approximate G6P/F6P as a single node for convenience. All Group A carbon sources enter the metabolic network through G6P/F6P, while Group B carbon sources enter the metabolic network from different points after glycolysis. Glycerol enters from the upper part of glycolysis but not G6P/F6P, thus we classify glycerol as a quasi-Group A carbon source (see Supplementary Note 1.4 for details).

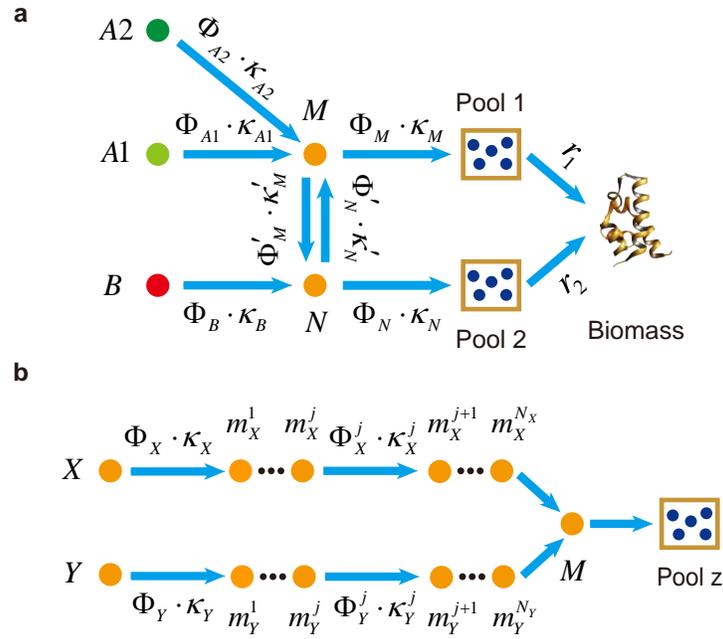

**Figure 2 | Topology of the metabolic network.** (**a**) Coarse-grained model of the metabolic network. Group A carbon sources merge to a common node $M$ before reaching precursor pools. Group B sources can supply some precursor pools from other routes. (**b**) Topology of the part of the metabolic network connecting two carbon sources to a precursor pool. The two carbon sources $X$ and $Y$ (from Group A and/or Group B) reach a common node $M$ through multiple intermediate nodes (metabolites) $m_X^j$ and $m_Y^j$ along their respective pathway, after which the flux is diverted to Pool $z$.

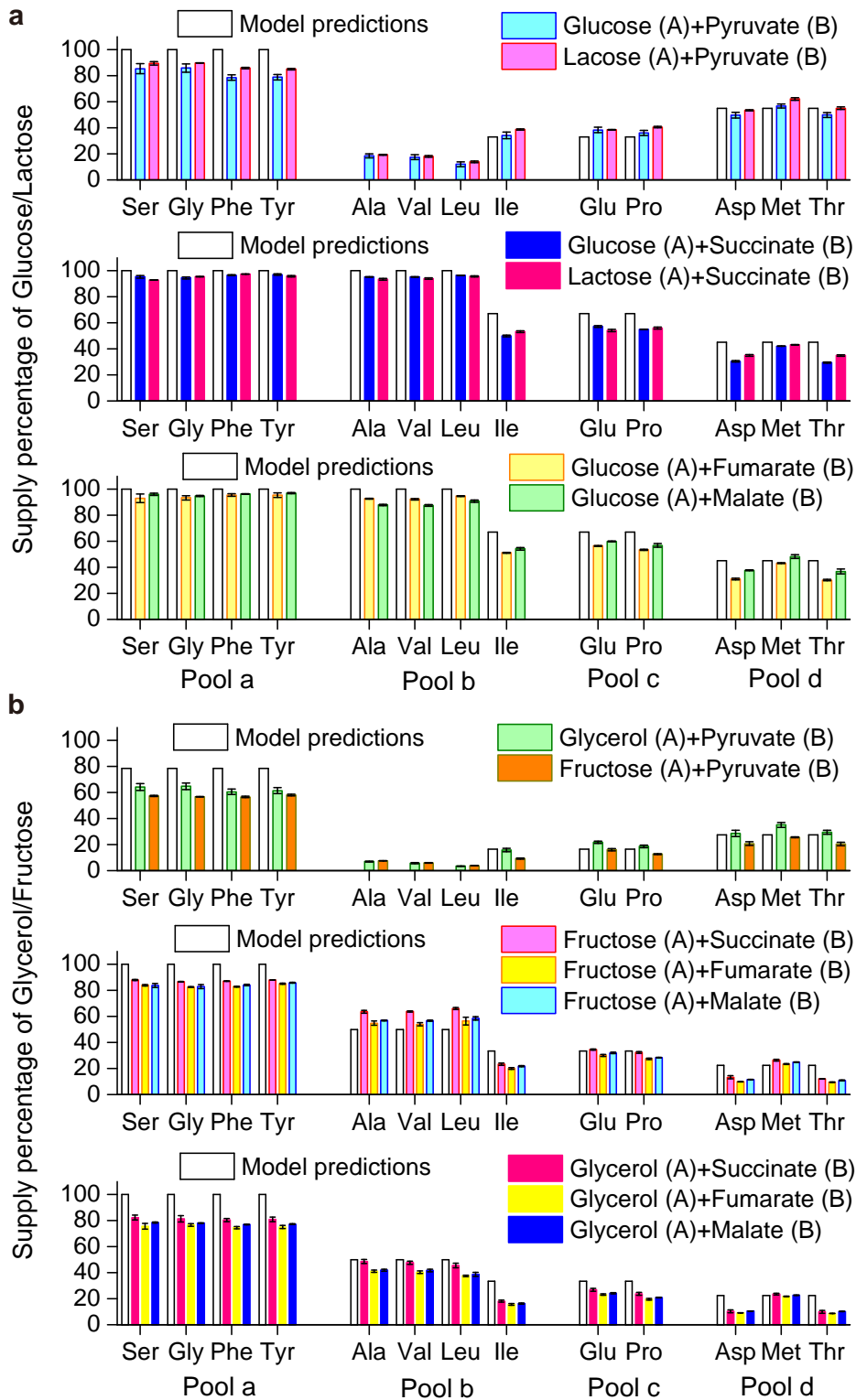

**Figure 3 | Suppliers of precursor pools in A+B cases.** Vertical axes are the percentages of the carbon atoms from the first of the two sources indicated. Model predictions (in

hollow bars, see Supplementary Table 4) are shown together with experimental results (in color bars). The source supplier of representative amino acids in Pools a-d was measured using $^{13}$C labeling (raw data from M-57 fragment; see Methods for details). Error bars represent standard deviations. Source data are provided as a Source Data file. **(a)** Glucose or lactose mixed with a Group B carbon source. **(b)** Fructose or glycerol mixed with a Group B carbon source.

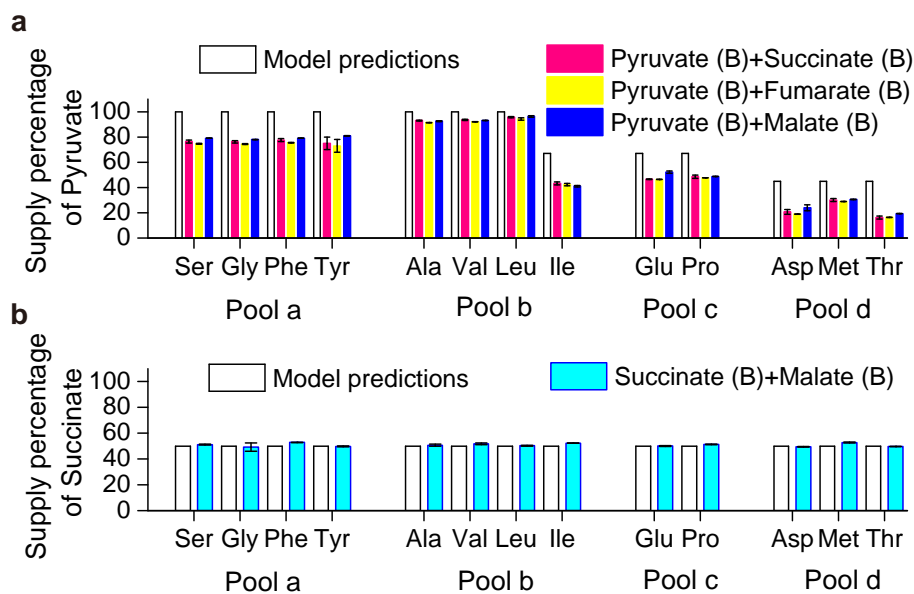

**Figure 4 | Suppliers of precursor pools in B+B cases.** Vertical axes are the percentages of the carbon atoms from the first of the two sources indicated. Model predictions (in hollow bars, see Supplementary Table 4) are shown together with experimental results (in color bars). The source supplier of representative amino acids in Pools a-d was measured using $^{13}$C labeling (raw data from M-57 fragment; see Methods for details). Error bars represent standard deviations. Source data are provided as a Source Data file. **(a)** Pyruvate mixed with another Group B carbon source. **(b)** Succinate mixed with malate. In this case, the branch efficiencies of the two sources are about the same.

# Growth strategy of microbes on mixed carbon sources

## Supplementary Information

Xin Wang et al.

## Supplementary Note 1. Model framework

### Supplementary Note 1.1 Optimization principles

We adopt an optimal protein allocation framework similar to that of previous studies[1-4]: Microbes optimize the efficiency of using enzymes through protein allocation. More specifically, microbes maximize the enzyme utilization efficiency $\varepsilon$:

$$\varepsilon \equiv \frac{J_{tot}}{\Phi_{tot}}, \qquad (1)$$

where $\Phi_{tot}$ is the total enzyme cost devoted to reactions and $J_{tot}$ is the total amount of carbon flux.

### Supplementary Note 1.2 Carbon flux, enzyme cost and substrate quality

Consider a biochemical reaction between substrate $S_i$ (with concentration $[S_i]$) and enzyme $E_i$ (with concentration $[E_i]$), assuming that $S_{i+1}$ is the product:

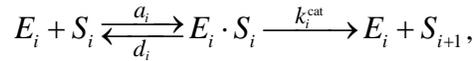

$$E_i + S_i \underset{d_i}{\overset{a_i}{\rightleftharpoons}} E_i \cdot S_i \xrightarrow{k_i^{cat}} E_i + S_{i+1},$$

where $a_i$, $d_i$ and $k_i^{cat}$ are chemical reaction parameters. The reaction rate (the carbon flux) $v_i$ follows Michaelis–Menten kinetics[5] (assuming that $[S_i] \gg [E_i \cdot S_i]$):

$$v_i = k_i^{cat} \cdot [E_i] \cdot \frac{[S_i]}{[S_i] + K_i}, \qquad (2)$$

where $K_i \equiv (d_i + k_i^{cat})/a_i$. The concentration of $E_i$ is defined as $[E_i] = \frac{N_{E_i}}{V_{cell}}$, where $V_{cell}$ is the cell volume and $N_{E_i}$ is the copy number of enzyme $E_i$ within the cell. The carbon flux of this reaction within the cell is:

$$J_i \equiv V_{cell} \cdot v_i. \qquad (3)$$

Denote the molecular weight (MW) of $E_i$ as $m_{E_i}$. To make the enzyme cost dimensionless, we define $m_0$ the MW unit, and $n_{E_i} \equiv m_{E_i}/m_0$ the cost of an $E_i$ molecule. Then the cost of all

$E_i$ molecules within a cell is

$$\Phi_i \equiv N_{E_i} \cdot n_{E_i} = V_{\text{cell}} \cdot [E_i] \cdot n_{E_i}. \tag{4}$$

The substrate quality of $S_i$ for enzyme $E_i$, $\kappa_i$, defined as the efficiency of using $E_i$, according to Supplementary Equation 1, follows:

$$\kappa_i \equiv \frac{J_i}{\Phi_i} = k_i \cdot \frac{[S_i]}{[S_i] + K_i}, \tag{5}$$

where

$$k_i \equiv k_i^{\text{cat}} / n_{E_i}. \tag{6}$$

Since $k_i$ and $K_i$ are constants, then $\kappa_i$ is a function of $[S_i]$.

**Supplementary Note 1.3 Biomass components and precursor pools**

For microbes, biomass consists of multiple components such as proteins, RNA, DNA, lipids, and glycogen, *etc.* (Fig. 1). Based on the topology of metabolic network, we classify the precursors of biomass components into seven pools. Specifically, each pool is named depending on its entrance point on the metabolic network: a1 (entering from G6P/F6P: precursors of RNA, DNA, Glycogen, Lipoglycans, Murein; His, Trp, Phe, Tyr), a2 (entering from GA3P: precursors of RNA, DNA; Trp, Phe, Tyr), a3 (entering from 3PG: Ser, Gly, Cys), a4 (entering from PEP: Trp, Phe, Tyr), b (entering from pyruvate/Acetyle-CoA: Lipids; Ala, Val, Leu, Ile), c (entering from α-Ketoglutarate: Glu, Gln, Pro, Arg) and d (entering from oxaloacetate: Asp, Asn, Met, Thr, Lys). In microbial growth, these seven pools draw roughly $r_{a1} = 24\%$, $r_{a2} = 14\%$, $r_{a3} = 5\%$, $r_{a4} = 5\%$, $r_b = 28\%$, $r_c = 12\%$ and $r_d = 12\%$ carbon flux, respectively[5-7]. Note that there are some overlapping components between Pools a1, a2, a3 and a4 owing to joint synthesis of precursors. For convenience, we lump sum Pools a1-a4 as Pool a and use the term precursor pools to denote both amino acid pools and the precursor pools for other components.

**Supplementary Note 1.4 Group A and Group B carbon sources**

Denote carbon sources entering the upper part of the glycolysis Group A and those joining at other parts of the metabolic network Group B (Fig. 1). Specifically, all Group A carbon sources join through G6P/F6P. Glycerol enters from the upper part of glycolysis but not G6P/F6P, thus we classify glycerol as a quasi-Group A carbon source. Group A carbon sources can be co-utilized with Group B sources, whereas substrates paired from Group A usually display diauxie. In most cases, glycerol follow all the traits of Group A sources, yet from the network topology, glycerol can be co-utilized with another Group A source of low concentration such as glucose and lactose (as recently observed in experiment[8]) under optimal conditions.

**Supplementary Note 1.5 Intermediate nodes**

In a real metabolic network (Fig. 1), there are multiple intermediate nodes in delivering carbon flux from carbon sources to precursor pools. We consider a simple case containing one carbon source $A1$ and one intermediate node, $M$ (Supplementary Fig. 1a). For $A1$, denote $E_{A1}$ as the catabolic enzyme (with enzyme cost $\Phi_{A1}$), the substrate concentration is $[A1]$ and the substrate quality is $\kappa_{A1} = k_{A1} \cdot \frac{[A1]}{[A1] + K_{A1}}$, where $k_{A1}$ and $K_{A1}$ are constants (similar to that of Supplementary Equation 5). For node $M$, $E_M$ denotes the catabolic enzyme (with enzyme cost $\Phi_M$), and $S_M$ the substrate (with concentration $[S_M]$ and substrate quality $\kappa_M$), where $\kappa_M = k_M \cdot \frac{[S_M]}{[S_M] + K_M}$. Here, $J_{tot} = \Phi_{A1} \cdot \kappa_{A1} = \Phi_M \cdot \kappa_M$, while $\Phi_{tot} = \Phi_{A1} + \Phi_M$. Then, $\Phi_{tot} = J_{tot}\left(1/\kappa_{A1} + 1/\kappa_M\right)$, and

$$\varepsilon = \frac{1}{1/\kappa_{A1} + 1/\kappa_M}. \tag{7}$$

$\varepsilon$ is clearly a monotonic function of $\kappa_M$ for a given nutrient condition (thus a given value of $\kappa_{A1}$). Combined with Supplementary Equation 5, it is clear that $\varepsilon$ is maximized when $[S_M]$ is nearly saturated:

$$\frac{[S_M]}{[S_M] + K_M} \approx 1, \tag{8}$$

where $K_M$ is the Michaelis–Menten constant, thus $\kappa_M \approx k_M$ (see Supplementary Equation 5). The real situations could be much more complicated (see Supplementary Notes 5-6 for representative cases), the substrate concentration of the intermediate nodes may be not saturated owning to other constraints. Strikingly, recent studies[9, 10] reported that at least in *E.coli*, metabolite concentration exceeds $K_M$ for most substrate-enzyme pairs. i.e. $[S_M] > K_M$, which implies $\kappa_M \approx k_M$.

**Supplementary Note 2. The origin of diauxie**

**Supplementary Note 2.1 Simplest model of diauxie**

In the simplest model for diauxie (Supplementary Fig. 1b), the carbon fluxes from substrates $A1$ and $A2$ infuse separately into the precursor pools. $\kappa_{Ai}$ ($i=1, 2$) is the substrate quality of $Ai$ ($i=1, 2$), while $E_{Ai}$ (with enzyme cost $\Phi_{Ai}$, $i=1, 2$) is the carrier enzyme for $Ai$. Here, $J_{tot} = \Phi_{A1} \cdot \kappa_{A1} + \Phi_{A2} \cdot \kappa_{A2}$, while $\Phi_{tot} = \Phi_{A1} + \Phi_{A2}$. Then,

$$\varepsilon = \frac{\Phi_{A1} \cdot \kappa_{A1} + \Phi_{A2} \cdot \kappa_{A2}}{\Phi_{A1} + \Phi_{A2}} = \kappa_{A1} - \frac{\Phi_{A2} \cdot (\kappa_{A1} - \kappa_{A2})}{\Phi_{A1} + \Phi_{A2}} = \kappa_{A2} - \frac{\Phi_{A1} \cdot (\kappa_{A2} - \kappa_{A1})}{\Phi_{A1} + \Phi_{A2}}. \qquad (9)$$

If $\kappa_{A1} > \kappa_{A2}$, then $\varepsilon = \kappa_{A1} - \frac{\Phi_{A2} \cdot (\kappa_{A1} - \kappa_{A2})}{\Phi_{A1} + \Phi_{A2}} \leq \kappa_{A1}$, with $\Phi_{A2} = 0$ the optimal point; if $\kappa_{A1} < \kappa_{A2}$, then $\varepsilon = \kappa_{A2} - \frac{\Phi_{A1} \cdot (\kappa_{A2} - \kappa_{A1})}{\Phi_{A1} + \Phi_{A2}} \leq \kappa_{A2}$, with $\Phi_{A1} = 0$ the optimal point. In either case, cells will only use the preferable carbon source, which corresponds to the case of diauxie.

**Supplementary Note 2.2 Branch efficiency**

In real cases, multiple intermediate nodes deliver carbon flux before branches converge to a common node. To take into account the cost of the intermediate enzymes, consider a model depicted in Fig. 2b. Here, carbon source $X$ mixed with carbon source $Y$, where $X$ and $Y$ can come from either Group A or Group B. Supposing that there are $N_X$ intermediate nodes ($m_X^j$, $j = 1 \sim N_X$) specifically for $X$ and $N_Y$ intermediate nodes ($m_Y^j$, $j = 1 \sim N_Y$) specifically for $Y$, with carbon fluxes from $X$ and $Y$ merging at node $M$. $M$ can be different for different combinations of $X$ and $Y$. $\kappa_X^j$, $\kappa_Y^j$ and $\kappa_M$ are the substrate quality of nodes $m_X^j$, $m_Y^j$ and $M$, respectively, and $E_X^j$, $E_Y^j$ and $E_M$ the corresponding carrier enzymes (with enzyme cost $\Phi_X^j$, $\Phi_Y^j$ and $\Phi_M$, respectively). Here, $J_{tot} = \Phi_X \cdot \kappa_X + \Phi_Y \cdot \kappa_Y$, with $\Phi_X \cdot \kappa_X = \Phi_X^j \cdot \kappa_X^j$ $(j=1 \sim N_X)$ and $\Phi_Y \cdot \kappa_Y = \Phi_Y^j \cdot \kappa_Y^j$ $(j=1 \sim N_Y)$, while

$$\Phi_{tot} = \Phi_X + \sum_{j=1}^{N_X} \Phi_X^j + \Phi_Y + \sum_{j=1}^{N_Y} \Phi_Y^j.$$ Then

$$\varepsilon = \frac{\Phi_X \cdot \kappa_X + \Phi_Y \cdot \kappa_Y}{\Phi_X + \sum_{j=1}^{N_X} \Phi_X^j + \Phi_Y + \sum_{j=1}^{N_Y} \Phi_Y^j} = \frac{\Phi_{X \to M}^b \cdot \varepsilon_{X \to M} + \Phi_{Y \to M}^b \cdot \varepsilon_{Y \to M}}{\Phi_{X \to M}^b + \Phi_{Y \to M}^b}, \quad (10)$$

where $\Phi_{X \to M}^b \equiv \Phi_X + \sum_{j=1}^{N_X} \Phi_X^j$ and $\Phi_{Y \to M}^b \equiv \Phi_Y + \sum_{j=1}^{N_Y} \Phi_Y^j$, while $\varepsilon_{X \to M}$ and $\varepsilon_{Y \to M}$ are defined as the branch efficiency of $X$ and $Y$ to convergent node $M$, respectively:

$$\varepsilon_{X \to M} = \frac{1}{1/\kappa_X + \sum_j^{N_X} 1/\kappa_X^j}, \varepsilon_{Y \to M} = \frac{1}{1/\kappa_Y + \sum_j^{N_Y} 1/\kappa_Y^j}. \quad (11)$$

Compare Supplementary Equation 10 with Supplementary Equation 9, it is clear that the supplier of node $M$ depends on the value of $\varepsilon_{X \to M}$ and $\varepsilon_{Y \to M}$. For $\varepsilon_{X \to M} > \varepsilon_{Y \to M}$, with $\Phi_{Y \to M}^b = 0$ the optimal point; For $\varepsilon_{X \to M} < \varepsilon_{Y \to M}$, with $\Phi_{X \to M}^b = 0$ the optimal point. Only the carbon source with higher branch efficiency is utilized to supply the convergent node $M$.

In the estimation of branch efficiency, $\kappa_X^j \approx k_X^j$ $(j=1 \sim N_X)$ and $\kappa_Y^j \approx k_Y^j$ $(j=1 \sim N_Y)$ (see Supplementary Equation 8 and Supplementary Note 1.5). $k_X^j$ and $k_Y^j$ are constants (see Supplementary Equation 6). A special case is that when $\varepsilon_{X \to M} \approx \varepsilon_{Y \to M}$, both branches are equally efficient. From optimality aspect, $X$ and $Y$ can supply node $M$ with any ratio, while from probability aspect, $X$ and $Y$ should supply node $M$ half and half. Rigorously, the application of branch efficiency in determining optimal supplier relies on that $\varepsilon_{X \to M}$ and $\varepsilon_{Y \to M}$ independent of $\Phi_{X \to M}^b$, $\Phi_{Y \to M}^b$ (see exceptional cases in Supplementary Note 6.5).

When $X$ and $Y$ both come from Group A, from the topology of metabolic network (Fig. 1 or coarse grained version Fig. 2a), the convergent node $M$ resides upstream to all precursor pools (Supplementary Fig. 1c). For example, for $X$ =glucose and $Y$ =galactose, $M$ = Glucose 6-phosphate (G6P); for the same $X$ but $Y$ =fructose, $M$ = fructose 6-phosphate (F6P). This leads to the result that Group A carbon sources are not utilized simultaneously.

**Supplementary Note 2.3 Decision line**

When we say $A1$ is a better carbon source than $A2$, we mean that at saturated concentrations of $A1$ and $A2$, the branch efficiency $\varepsilon_{A1 \to M} > \varepsilon_{A2 \to M}$ (Supplementary Fig. 1c). However, Supplementary Equation 5 and Supplementary Equation 11 indicate that branch efficiency depends on nutrient concentration $[Ai]$, via the substrate quality $\kappa_{Ai} = k_{Ai} \cdot \frac{[Ai]}{[Ai] + K_{Ai}}$. $\varepsilon_{Ai}$ is a monotonic function of $[Ai]$. Thus a better sugar at low concentration may not be as preferable as a worse sugar at high concentration. As shown in Supplementary Fig. 1c, assuming that there are $N_{A1}$ intermediate nodes ($m_{A1}^j$, $j = 1 \sim N_{A1}$) specifically for $A1$ and $N_{A2}$ intermediate nodes ($m_{A2}^j$, $j = 1 \sim N_{A2}$) specifically for $A2$. $\kappa_{A1}^j \approx k_{A1}^j$ and $\kappa_{A2}^j \approx k_{A2}^j$ (see Supplementary Equation 5 and Supplementary Note 1.5) are the substrate quality of $m_{A1}^j$ and $m_{A2}^j$, respectively. Ideally, the decision line to switch the sugar source and thus to turn on $A2$ carrier genes is at

$$\varepsilon_{A1 \to M}([A1]) = \varepsilon_{A2 \to M}([A2]). \tag{12}$$

which is

$$1/\kappa_{A1} = 1/\kappa_{A2} + \left( \sum_j^{N_{A2}} 1/k_{A2}^j - \sum_j^{N_{A1}} 1/k_{A1}^j \right), \tag{13}$$

Substituting $\kappa_{Ai} = k_{Ai} \cdot \frac{[Ai]}{[Ai] + K_{Ai}}$, Supplementary Equation 12 is reduced to

$$[A1] = \frac{\delta \cdot [A2]}{\Delta + [A2]}, \tag{14}$$

where $\delta = \frac{K_{A1}}{c_{A1}^{A2} \cdot k_{A1}}$ and $\Delta = \frac{K_{A2}}{c_{A1}^{A2} \cdot k_{A2}}$, and $c_{A1}^{A2}$ is defined as $c_{A1}^{A2} \equiv \frac{1}{\varepsilon_{A2}^{max}} - \frac{1}{\varepsilon_{A1}^{max}}$, with

$\varepsilon_{Ai}^{max} = \frac{1}{1/k_{Ai} + \sum_{j}^{N_{Ai}} 1/k_{Ai}^{j}}$ the maximum efficiency of the nutrient $Ai$ (at saturating nutrient concentration). $k_{Ai}$ is defined according to Supplementary Equation 6. Since $\varepsilon_{A1}^{max} > \varepsilon_{A2}^{max}$, thus $c_{A1}^{A2}$, $\delta$ and $\Delta$ are all positive constants. When $[A2]$ is small ($[A2] \ll \Delta$), the decision line (Supplementary Equation 14) is reduced to $[A1] = \frac{\delta}{\Delta} \cdot [A2]$. That is, the decision line depends only on the ratio of the two nutrients: the cell will sense not the absolute concentration of $[A1]$ and $[A2]$, but their ratio, to make the decision. Ratio sensing was recently observed in the budding yeast *Saccharomyces cerevisiae* cultured in glucose-galactose mixed medium[11]. The measured turning point agrees remarkably well with Supplementary Equation 14 (see Supplementary Fig. 2).

Meanwhile, the mechanism of ratio sensing demands resources. It could well be that the microbe cares only about the most frequently encountered (or the most important) combinations of nutrients and would not invest resources to ratio sense the others.

**Supplementary Note 3. The reason for co-utilization**

The topologies of the metabolic network with one carbon source from Group A and the other from Group B are shown in Supplementary Fig. 3a-c. Note a Group B carbon source has the possibility to be co-utilized with another source from Group B. A topology of this type is shown in Supplementary Fig. 3d. These diagrams can be analyzed with the coarse grained model of Supplementary Fig. 1d, which has two precursor pools. In synthesizing biomass, portion $r_1$ of carbon flux comes from Pool 1 and $r_2$ from Pool 2. All intermediate nodes are lump summed into two intermediate nodes: $M$ and $N$, which can convert to each other with the help of respective enzymes: $E_M'$ and $E_N'$. $\kappa_A$, $\kappa_B$, $\kappa_M$, $\kappa_N$, $\kappa_M'$ and $\kappa_N'$ represent substrate quality, while $\Phi_A$, $\Phi_B$, $\Phi_M$, $\Phi_N$, $\Phi'_M$ and $\Phi'_N$ denote protein cost of carrier enzymes.

Here, $\Phi_{tot} = \Phi_A + \Phi_B + \Phi_M + \Phi_N + \Phi'_M + \Phi'_N$, $J_{tot} = \Phi_M \cdot \kappa_M + \Phi_N \cdot \kappa_N$ with constraints: $\Phi_M \cdot \kappa_M + \Phi'_M \cdot \kappa'_M = \Phi_A \cdot \kappa_A + \Phi'_N \cdot \Phi'_N$, $\Phi_N \cdot \kappa_N + \Phi'_N \cdot \kappa'_N = \Phi_B \cdot \kappa_B + \Phi'_M \cdot \kappa'_M$ and $\dfrac{\Phi_M \cdot \kappa_M}{\Phi_N \cdot \kappa_N} = \dfrac{r_1}{r_2}$. To maximize enzyme utilization efficiency $\varepsilon = \dfrac{J_{tot}}{\Phi_{tot}}$, it is equivalent to maximize the enzyme utilization efficiency of every precursor pools (Pool 1 and Pool 2 in Supplementary Fig. 1d), where we can apply branch efficiency analysis (see Supplementary Note 2.2 for details) -- only the carbon source with the highest branch efficiency will be employed to supply the convergent node. At the convergent node $M$: $\varepsilon_{A \to M} = \dfrac{1}{1/\kappa_A}$, $\varepsilon_{B \to M} = \dfrac{1}{1/\kappa_B + 1/\kappa'_N}$.

The nutrient supplier of Pool 1 is then determined by

$$\text{Pool 1 is supplied by } \begin{cases} A, \text{ if } \varepsilon_{A \to M} > \varepsilon_{B \to M} \\ B, \text{ if } \varepsilon_{A \to M} < \varepsilon_{B \to M} \end{cases}. \quad (15)$$

At the intersection node $N$: $\varepsilon_{A \to N} = \dfrac{1}{1/\kappa_A + 1/\kappa'_M}$, $\varepsilon_{B \to N} = \dfrac{1}{1/\kappa_B}$. The provider of Pool 2 is then determined according to

$$\text{Pool 2 is supplied by } \begin{cases} A, \text{ if } \varepsilon_{A \to N} > \varepsilon_{B \to N} \\ B, \text{ if } \varepsilon_{A \to N} < \varepsilon_{B \to N} \end{cases}. \quad (16)$$

If

$$1/\kappa_B - 1/\kappa'_M < 1/\kappa_A < 1/\kappa_B + 1/\kappa'_N, \quad (17)$$

then $A$ supplies Pool 1 and $B$ provides Pool 2, both substrates are co-utilized. In this case, the enzyme utilization efficiency $\varepsilon$ of the mixed medium $A+B$ is (see Supplementary Equation 1)

$$\varepsilon_{A+B} = \dfrac{1}{r_1/\kappa_A + r_1/\kappa_M + r_2/\kappa_B + r_2/\kappa_N}, \quad (18)$$

while the enzyme utilization efficiency $\varepsilon$ of a single substrate medium like $A1$ is

$$\varepsilon_A = \dfrac{1}{1/\kappa_A + r_1/\kappa_M + r_2/\kappa_N + r_2/\kappa'_M}. \quad (19)$$

In the real case, $1/\kappa'_M$ delegates the summation of intermediate node terms between *M* and *N*, which is often quite large (Supplementary Table 2). As a result, $\varepsilon_{A+B}$ can be significantly greater than $\varepsilon_A$, meaning that co-utilization is commonly the optimal choice when a Group A carbon source mixed with a Group B carbon source.

**Supplementary Note 4. Pool suppliers in the case of co-utilization**

**Supplementary Note 4.1 The original suppliers of pools in the cases of co-utilization**

The original pool suppliers are determined by the branch efficiencies. We collected the available biochemical parameters from published data (Supplementary Table 1) to estimate the branch efficiencies from carbon sources to precursor pools in *E. coli* (Supplementary Table 2).

In our estimation, for convenience, we only consider carbon sources (e.g. glucose, lactose, pyruvate etc.) with saturated concentrations. For intermediate metabolites, since $[S_M] > K_M$ is valid for most substrate-enzyme pairs in *E.coli*[9, 10], we estimate the substrate quality $\kappa_i \approx k_i = k_i^{\text{cat}}/n_{E_i}$ (see Supplementary Equation 6 and Supplementary Note 1.5). $k_i^{\text{cat}}$ is the turnover number of the enzyme $E_i$, $n_{E_i} = m_{E_i}/m_0$ (see Supplementary Note 1.2) with $m_{E_i}$ the MW of $E_i$ and $m_0$ a MW unit. For convenience, we set $m_0 = 100$ kDa and then can obtain branch efficiency from carbon sources to the counterpart of node *M* and *N* (F6P, GA3P, 3PG, PEP, pyruvate or oxaloacetate) of Pools a1-a4, b and d (Supplementary Table 2). Using branch efficiency analysis (see Supplementary Note 2.2), we get the original suppliers of each pool in different combinations of co-utilization (Supplementary Table 3). Take the case of glucose-pyruvate co-utilization for example (Supplementary Table 3 and Supplementary Fig. 4a), the original supplier of Pools a1-a4 and d is glucose, while the original supplier of Pool b is pyruvate. Due to converged flux, the original supplier of Pool c is not set by the branch efficiency from carbon sources to Pool c, but rather both the suppliers of Pool b (pyruvate) and Pool d (glucose) (see Supplementary Note 4.2 for details). It is worth noting that owing to energy production in the TCA cycle, the pool suppliers in practice can be different from its original supplier (see Supplementary Notes 4.3-4.5 for details).

**Supplementary Note 4.2 Converged flux**

(a) Pool c

Pool c is supplied by joint fluxes from pyruvate and oxaloacetate (Supplementary Fig. 4b, pyruvate→ Acetyl-CoA, Acetyl-CoA + oxaloacetate → citrate), which are $M$ and $N$ node counterpart of Pools b and d, respectively, thus Pool c suppliers are both that of Pools b and d, with 2/5 of carbons supplied from Pools b and 3/5 of carbons supplied from Pools d.

(b) Isoleucine

Isoleucine is synthesized through the following reactions[5]:

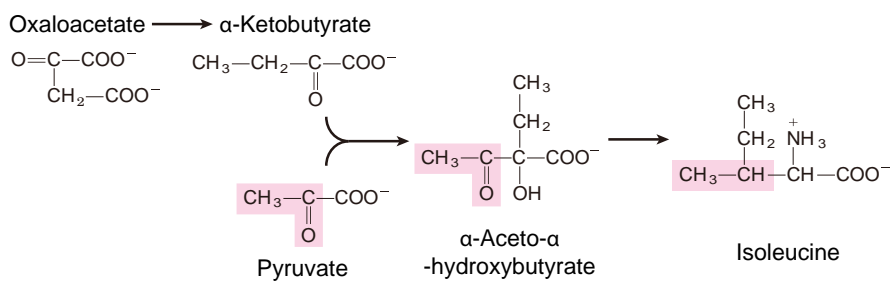

with 2/5 of carbons supplied from Pools b and 3/5 of carbons supplied from Pools d.

(c) Tryptophan, Tyrosine & Phenylalanine

Tryptophan, tyrosine and phenylalanine are synthesized through the following reactions[5]:

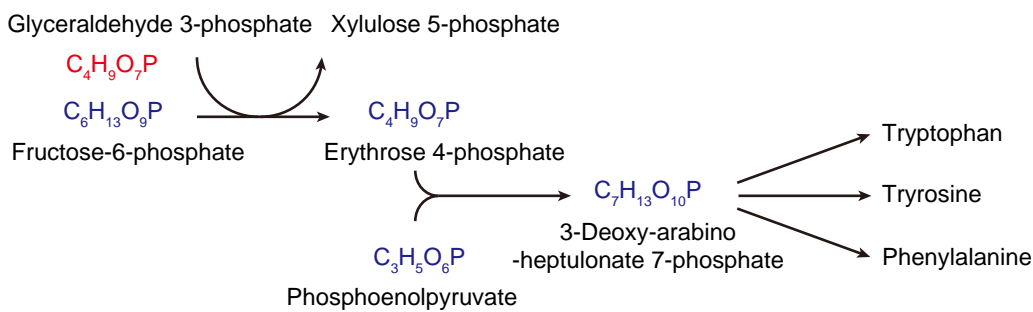

with roughly 3/7 of carbons supplied from phosphoenolpyruvate (PEP) and 4/7 of carbons supplied from fructose-6-phosphate (F-6-P).

**Supplementary Note 4.3 Energy production**

In microbial growth, a considerable amount of carbon sources needs to be allocated for energy production. Adenosine triphosphate (ATP), the molecular unit of energy currency, facilitates intracellular energy transfer. Taking the growth of *E. coli* in glucose as an example, it is estimated that $1\text{-}2\times10^9$ glucose molecules ($7\times10^9$ carbon atoms/cell, BNID 103010) are required to build the biomass of a new cell[12], whereas the amount of carbon sources required for energy production varies depending on the respiration type. For typical aerobic respiration, the energetic requirement for a new cell is estimated to be $3-6\times 10^8$ glucose molecules ($1.2$-$1.7\times 10^{10}$ ATP/cell, BNID 101981, 101983) on top of the molecules needed for the biomass[12-14]. For anaerobic respiration, $3-6\times 10^9$ glucose molecules[12] are estimated to be required for energy production of a new cell. For aerobic microbial growth in mixed carbon sources with saturated carbon source concentrations, the ratio of energy/biomass allocation is estimated to be 20%-50%[12-14], and we denote this ratio as $r_{E/M}$.

**Supplementary Note 4.4 Pool suppliers influenced by the TCA cycle**

In TCA cycle, oxaloacetate goes back to itself if reactions flow through the whole cycle (with production of $CO_2$ and ATPs). Consider a newly synthesized oxaloacetate molecule coming from the original supplier of Pool d before the TCA cycle, as depicted in Supplementary Fig. 4b, after 1 round of TCA cycle, half of the carbon atoms in oxaloacetate are replaced by those coming from pyruvate (supplier of Pool b), which means that in practice the supplier of Pool d would be a combination of the original supplier of Pool d and the supplier of Pool b. For clarity, we denote the original supplier of Pool d as $d_S^O$, the original supplier of Pool b as $b_S^O$ and the supplier of Pool d in practice as $d_S^I$. Clearly, $d_S^I$ is a combination of $b_S^O$ and $d_S^O$, and we assume that $\xi$ fraction of carbon atoms in $d_S^I$ coming from $d_S^O$, with the rest $1-\xi$ carbon atoms coming from $b_S^O$.

To quantify the influence of TCA cycle on $\xi$, we consider microbes of exponential growth at certain growth rate, with the TCA cycle depicted in Supplementary Fig. 4c. Since biomass production and energy production are continuous, we assume the total stoichiometry of carbon flux at oxaloacetate is 1 per unit time $\tau$. $r_d'$ stoichiometry of the flux flows to Pool d, while $1-r_d'$ stoichiometry of flux flows to citrate. Owing to the stoichiometry of reactions (pyruvate→ Acetyl-CoA and Acetyl-CoA + oxaloacetate → citrate), $1-r_d'$ stoichiometry of carbon flux would come from citrate per $\tau$. Meanwhile, $r_c'$ stoichiometry of the flux flows to Pool c, with the rest $1-r_c'-r_d'$ stoichiometry of carbon flux flows back to oxaloacetate. To keep sustainable microbial growth, $r_c' + r_d'$ stoichiometry of carbon flux per $\tau$ would join oxaloacetate from $d_S^O$ so that the stoichiometry of flux at oxaloacetate can be 1 (our assumption from beginning). For microbial growth at fixed growth rate, the system can be treated as non-equilibrium steady state[15], then $\xi$ is under the constraint of the following equation:

$$\xi = \left(r_c' + r_d'\right) + \left(1 - r_c' - r_d'\right)\frac{\xi}{2}, \tag{20}$$

which means that the value of $\xi$ should be the same after 1 round of TCA cycle. The 1/2 in Supplementary Equation 20 derives from the fact that half of the carbon atoms in oxaloacetate have been replaced by that of pyruvate through a TCA cycle. We can solve Supplementary Equation 20 and get:

$$\xi = 2\frac{r_c' + r_d'}{1 + r_c' + r_d'}. \tag{21}$$

In fact, we can estimate the values of $r_c'$ and $r_d'$ using $r_c$, $r_d$ and $r_{E/M}$. Note that $r_c'$ and $r_d'$ denote the stoichiometry ratios of carbon flux. At the point of oxaloacetate before a TCA cycle, 1 stoichiometry flux of oxaloacetate (with 4 carbon atoms per molecule) corresponds to 4 fluxes of carbon atoms. Then, $r_d'$ stoichiometry of oxaloacetate means $4r_d'$ of carbon atoms flow to Pool d. Because of the stoichiometry of reaction (Acetyl-CoA + oxaloacetate →citrate), another $2\left(1 - r_d'\right)$ of carbon atoms join from Acetyl-CoA (Supplementary Fig. 4b-c), this results in $6\left(1 - r_d'\right)$ of carbon atoms at citrate. Later on, owing to carbon leakage in the forms of carbon dioxide ($CO_2$), $5r_c'$ of carbon atoms flows to Pool c through α-ketoglutarate (with 5 carbon atoms per molecule), with finally $4\left(1 - r_c' - r_d'\right)$ of carbon atoms flowing back to oxaloacetate. Clearly, the carbon atoms flowing to Pools c & d are $5r_c'$ and $4r_d'$ per $\tau$, respectively. Meanwhile, the carbon flux allocated for energy production can be estimated from the production of $CO_2$. As shown in Supplementary Fig. 4b-c, $CO_2$ is generated of $1 - r_d'$ in reaction pyruvate→Acetyl-CoA; $1 - r_d'$ in reaction of isocitrate→α-ketoglutarate, and $1 - r_c' - r_d'$ in reaction

α-ketoglutarate→Succinyl-CoA, with $3 - r'_c - 3r'_d$ per $\tau$ in total. We roughly estimate this to be the energy allocation for generating ATPs. Then,

$$\frac{r_{E/M}}{3 - r'_c - 3r'_d} = \frac{r_c}{5r'_c} = \frac{r_d}{4r'_d}, \quad (22)$$

thus

$$\begin{cases} r'_c = \dfrac{3r_c}{5r_{E/M} + r_c + 3.75r_d} \\ r'_d = \dfrac{3.75r_d}{5r_{E/M} + r_c + 3.75r_d} \end{cases}. \quad (23)$$

Substitute Supplementary Equation 23 into Supplementary Equation 21, we get:

$$\xi = \frac{6r_c + 7.5r_d}{5r_{E/M} + 4r_c + 7.5r_d}. \quad (24)$$

Since $r_c \approx 0.12$, $r_d \approx 0.12$, $0.2 \leq r_{E/M} \leq 0.5$ (Supplementary Notes 1.3 & 4.3), the value of $\xi$ can be estimated from Supplementary Equation 24:

$$0.42 \leq \xi \leq 0.68. \quad (25)$$

Combined with the experiment data we obtained, we roughly estimate $\xi$ to be 0.55 for all the aerobic growth with saturated carbon source concentrations.

**Supplementary Note 4.5 Pool suppliers in practice for optimal growth**

To summarize, for optimal growth of *E.coli*, $\xi$ (≈55%) of the carbons in Pool d (Pool d supplier in practice $d_S^I$) come from its original supplier $d_S^O$ (determined by its branch efficiency), $1-\xi$ (≈45%) of the carbons in Pool d come from the supplier of Pool B $b_S^O$. Meanwhile, Pool c is supplied by joining fluxes from pyruvate (2/5) and oxaloacetate (3/5). Consequently, $0.6\xi$ (≈33%) of the carbons in Pool c come from $d_S^O$ while $1-0.6\xi$ (≈67%) of the carbons in Pool c come from $b_S^O$. Model predictions of the pool suppliers for some combinations of mixtures are listed in Supplementary Table 4.

**Supplementary Note 4.6 Note on pathways we have considered**

Besides the metabolite pathways listed in Fig. 1 (with enzyme parameters shown in Supplementary Table 1), we have considered other pathways such as Entner-Doudoroff pathway and Glyoxylate cycle[5], yet the inclusion of these pathways do no change the model predictions of carbon source suppliers for the combinations of mixtures we have considered.

**Supplementary Note 4.7 Spontaneous decarboxylation of oxaloacetate in solution**

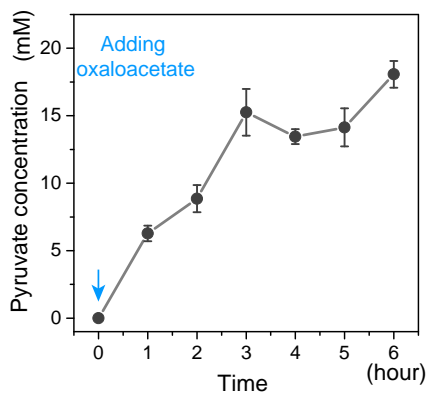

Figure. Oxaloacetate spontaneously decarboxylates into pyruvate in solution. Oxaloacetate powders were added into solution at time 0. Each dot is an average over three independent experiments, error bars represent standard deviations. Source data are provided as a Source Data file.

Oxaloacetate can spontaneously decompose to pyruvate and $CO_2$ when added into solution[16]. We confirmed this result with our experiment (shown above) and found that the quick decarboxylation of oxaloacetate would result in high concentration of pyruvate. This essentially changes the types of carbon sources in the culturing medium. As a result, we do not predict combinations including oxaloacetate.

**Supplementary Note 4.8 Enzymes of gluconeogenesis**

For optimal growth, there should be no gluconeogenesis enzymes when there is carbon flux of glycolysis. In reality, microbes need to dealing with the frequently varying environments. Empirically, it was found in *E.coli* that microbes reserve a portion of gluconeogenesis enzymes when using Group A carbon sources, with the enzymes expression level anti-correlated with the carbon fluxes down through the glycolysis[17]. This may enable microbes to balance growth and prepare for potential changing environments[17].

**Supplementary Note 4.9 Predictions compared with experiments**

The comparison of experimental results with predictions of the in-practice pool suppliers (Supplementary Table 4) is shown in Figs. 3-4 & Supplementary Figs. 5-6. For all these combinations, the experimental results quantitatively agree with the model predictions. The slight differences between model predictions and experimental results in Pool a (combining Pool a1-a4) supply might be due to the fact that microbes reserve a portion of gluconeogenesis enzymes to prepare for potential changing environment[17].

In our model predictions, all parameters were collected or estimated from published literatures (Supplementary Table 1). For some transporters with no direct experimental data, we estimated the order of magnitude for their parameters[12] (labeled as "Estimated" in Supplementary Table 1). Admittedly the reported turnover numbers (or $k_{cat}$ values) of enzymes are likely to be associated with errors (e.g. measurement errors and *in vitro* versus *in vivo* errors), and the estimated parameters for transporters probably involve even large errors. However, it is still rather nontrivial to observe the consistency between model predictions (Supplementary Table 4) and experimental data (Figs. 3-4 & Supplementary Figs. 5-6) as we explain it below.

In Supplementary Tables 6 and 7, we list all possible patterns of carbon source distribution for both of the original pool suppliers (Supplementary Table 6) and of the in-practice pool suppliers (Supplementary Table 7) of *E. coli*, respectively, where we have considered all possibilities for the choice of the biochemical parameters ($k_{cat}$ can be any positive values for all enzymes involved). For each combination of carbon sources, there are only a handful of possible patterns (Supplementary Tables 6 and 7), since there can only be one-zero or half-half (when the branch efficiencies of two substrates are roughly the same for a precursor pool) supply pattern for the original supplier of a precursor pool and the branch efficiency of every carbon source decreases when incorporating more intermediate nodes. Evidently, all possible patterns are very distinct from each other (quantized), and actually there is no way to freely fit even a single pattern in any mixture of two carbon sources by tuning parameters. To illustrate this point, we focus on possible patterns of the in-practice pool suppliers (Supplementary Table 7) that are directly comparable to experimental results.

In the case that a Group A carbon source mixed with B1 (pyruvate) (A+B1 in Supplementary Table 7), there are two possible patterns of diauxie (DX Nos. 1-2) and 13 possible patterns of co-utilization (CoU Nos. 1-13). However, only three patterns (CoU No. 1, CoU No. 8 and CoU No. 9 in A+B1) are qualitatively similar to experimental results (A+B1); other patterns are very different in at least one of the pools (Pool a-d). Among the three similar patterns, CoU No. 1 (in A+B1) is the pattern predicted for Glucose/Lactose + B1 (Supplementary Table 4); CoU No. 9 (in A+B1) is the pattern predicted for Fructose/Glycerol + B1 (Supplementary Table 4); CoU No. 8 (in A+B1) is very similar to CoU No. 9 (in A+B1), yet one necessity for CoU No. 8 is that pyruvate is more efficient to supply oxaloacetate (entry point of Pool d) via malate than the other route via PEP (entry point of Pool a4), which requires at least 10-fold increase in the $k_{cat}$ value (in the reaction: Pyruvate →Malate) from the nominal value based on published literatures (Supplementary Table 1). Obviously, there is no free space for parameter fitting, and this situation also holds for most mixed combinations.

In the case that a Group A carbon source mixed with B2 (Succinate/ Malate/ Fumarate) (A+B2 in Supplementary Table 7), there are two possible patterns of diauxie (DX Nos. 1-2) and 17 possible patterns of co-utilization (CoU Nos. 1-17). For Glucose/Lactose + B2, only two possible patterns are qualitatively similar with the experiment: CoU No. 1 and CoU No. 8 (in A+B2). Here CoU No. 1 (in A+B2) is the predicted pattern; CoU No. 8 (in A+B2) is actually possible when tuning the $k_{cat}$ values of B2 transporters (requires a reduction of roughly forty percent, in Supplementary Table 1). Thus, in this case, tuning parameters can facilitate a choice between CoU No. 1 and CoU No. 8 (in A+B2), yet there is no other choice for a qualitatively similar pattern. Quantitatively, CoU No. 8 (in A+B2) is distinguishable from CoU No. 1 (in A+B2) and different from the experimental pattern. For Fructose/Glycerol + B2, there are four patterns qualitatively similar with experiments: CoU No. 7, CoU No. 9, CoU No. 11 and CoU No. 13 (in A+B2). Here CoU No. 9 (in A+B2) is the predicted pattern. CoU No. 13 (in A+B2) is very similar to CoU No. 9 (in A+B2), yet CoU No. 13 requires more than 5-fold increase in the $k_{cat}$ value (in the reaction: oxaloacetate →PEP) from the nominal value based on published literatures (Supplementary Table 1). CoU No. 7 and CoU No. 11 (in A+B2) are quantitatively distinguishable from CoU No. 9 (in A+B2). Based on current parameters in Supplementary Table 1, A owns higher branch efficiency at pyruvate (entry point of Pool b) than that of oxaloacetate (entry point of Pool d), then both CoU No. 7 and CoU No. 11 (in A+B2) would require that B2 owns a higher branch efficiency at pyruvate than that of oxaloacetate, which means a minimum 30-fold increase in the $k_{cat}$ value (in the reaction: Malate →Pyruvate) from the nominal value based on published literatures (Supplementary Table 1). Meanwhile, to make A with a higher branch efficiency at oxaloacetate than that of pyruvate would require at least 6-fold increase in the $k_{cat}$ value (in the reaction: PEP →Oxaloacetate) from the nominal value based on published literatures (Supplementary Table 1). Furthermore, CoU No. 11 (in A+B2) additionally requires that the branch efficiencies from A and B2 equal to each other both at pyruvate and oxaloacetate via two independent equalities.

In the case that B1 (pyruvate) mixed with B2 (Succinate/ Malate/ Fumarate) (B1+B2 in Supplementary Table 7), two patterns are qualitatively similar with experiments: CoU No. 1 and CoU No. 3 (in B1+B2). Here CoU No. 1 (in B1+B2) is the predicted pattern; CoU No. 3 (in B1+B2) is also possible when tuning the $k_{cat}$ values of B2 transporters (requires a reduction of roughly forty percent, in Supplementary Table 1), and quantitatively it is distinguishable from CoU No. 1 (in B1+B2) and different from the experimental pattern.

Finally, in the case that Succinate mixed with Malate (Succinate + Malate in Supplementary Table 7), CoU No. 1 (Succinate + Malate) is the only co-utilized pattern and also the predicted pattern, which quantitatively agree with experimental results.

Summarily, for any mixture, only a very limited number (no more than 4) of possible patterns can qualitatively agree with experiments. While among these qualitatively similar patterns, if the associated errors of the $k_{cat}$ (in Supplementary Table 1) are less than 5-fold, tuning parameters can only facilitate two choices: one between CoU No. 1 and CoU No. 8 in A+B2 and the other between CoU No. 1and CoU No. 3 in B1+B2. Thus, tuning parameters cannot freely fit pool supply patterns. Consequently, the consistency between model predictions (Supplementary Table 4 or blue text lines in Supplementary Table 7) and experimental data in all combinations of mixtures considered demonstrates the usefulness of our theory.

Furthermore, besides the *E. coli* data from our experiments, published experiment data[18] in *Methylobacterium extorquens AM1* is qualitatively highly consistent with our model. Their carbon supply pattern (Figure 3 in the reference article[18]) is very similar to that of Supplementary Fig. 4a.

**Supplementary Note 5. Reversible reactions**

Reversible reactions are common in metabolic network (Fig. 1). To analyze the influence of this factor, we consider the scheme that

$$S_i \xleftrightarrow{E_i} S_{i+1},$$

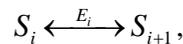

where $E_i$ is the enzyme catalyzing the reversible reaction between substrate $S_i$ and $S_{i+1}$. We can approximate the details of the reaction as follows:

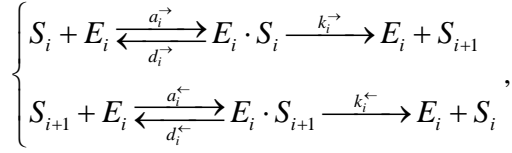

Here $a_i^\rightarrow$, $a_i^\leftarrow$, $d_i^\rightarrow$, $d_i^\leftarrow$, $k_i^\rightarrow$ and $k_i^\leftarrow$ are the chemical reaction parameters, the superscripts $\rightarrow$ and $\leftarrow$ stand for forward and reverse reactions, respectively. The net reaction rate $v_i$ from $S_i$ to $S_{i+1}$ follows the Michaelis-Menten equation of the reversible form[19, 20]:

$$v_i = \frac{k_i^\rightarrow [S_i]/K_i^\rightarrow - k_i^\leftarrow [S_{i+1}]/K_i^\leftarrow}{1+[S_i]/K_i^\rightarrow + [S_{i+1}]/K_i^\leftarrow}[E_i], \quad (26)$$

where $K_i^\rightarrow \equiv (d_i^\rightarrow + k_i^\rightarrow)/a_i^\rightarrow$, and $K_i^\leftarrow \equiv (d_i^\leftarrow + k_i^\leftarrow)/a_i^\leftarrow$. Then, the substrate quality of $S_i$:

$$\kappa_i = \frac{k_i^\rightarrow [S_i]/K_i^\rightarrow - k_i^\leftarrow [S_{i+1}]/K_i^\leftarrow}{1+[S_i]/K_i^\rightarrow + [S_{i+1}]/K_i^\leftarrow} < k_i^\rightarrow \quad . \quad \text{Note that when} \quad [S_i]/K_i^\rightarrow > [S_{i+1}]/K_i^\leftarrow > 1$$

(rigorously $[S_i]/K_i^\rightarrow \gg [S_{i+1}]/K_i^\leftarrow \gg 1$), $\varepsilon$ is maximized and $\kappa_i \approx k_i^\rightarrow$, which means that at optimal conditions analysis applies to irreversible cases is valid for reversible reactions.

One prediction of the reversible reaction analysis is that metabolites at the upper part of glycolysis (e.g. G6P, fructose 1,6-bisphosphatase (FBP)) owns a much higher concentration when bacteria cultured in Group A carbon sources (e.g. glucose) compared to that bacteria cultured in carbon sources entering from lower parts of glycolysis or TCA cycle. Recent studies[17] found that G6P and FBP have a much higher concentration when bacteria cultured with glucose than that shifting into acetate (entering from the bottom of glycolysis), which agree well with our reversible reaction analysis.

**Supplementary Note 5.1 Influence of the reversible reactions**

Note that optimal conditions operate at a non-equilibrium steady state that metabolic flux coming from external carbon sources working at maximum rate. Were the carbon flux drops for a while (e.g. bacteria take a short rest when consuming sugars), the reversible reaction between $S_i$ and $S_{i+1}$ would be quickly in equilibrium. This ideally can make the pools with reversible interconverting entry metabolites (counterpart of node *M* and *N*) be at similar carbon supply percentages. For the synthesis of biomass, Pools a2-a4 can be affected by this effect which makes the in-practice suppliers of Pools a2-a4 very similar.

## Supplementary Note 6. Metabolic regulations

### Supplementary Note 6.1 Enzyme concentration dependent reaction rate

Consider again enzyme reaction:

$$S_i + E_i \underset{d_i}{\overset{a_i}{\rightleftharpoons}} E_i \cdot S_i \overset{k_i^{cat}}{\longrightarrow} E_i + S_{i+1}.$$

The Michaelis-Menten enzyme kinetics $v_i = k_i^{cat} \frac{[S_i]}{[S_i]+K_i}[E_i]$ actually relies on the assumption that $[S_i] \gg [E_i \cdot S_i]$ [5], yet the precise form of the reaction rate is enzyme concentration dependent[21, 22]:

$$v_i = k_i^{cat} \frac{([S_i]+[E_i]+K_i)}{2}\left(1-\sqrt{1-\frac{4[S_i][E_i]}{([S_i]+[E_i]+K_i)^2}}\right) \approx k_i^{cat} \frac{[S_i][E_i]}{[E_i]+[S_i]+K_i}. \quad (27)$$

Here we apply approximation $\sqrt{1-x} \approx 1-x/2$ where $x \equiv \frac{4[S_i][E_i]}{([S_i]+[E_i]+K_i)^2} < 1$. When $\varepsilon$ is maximized, $[S_i] \gg K_i, [E_i]$, and $\kappa_i \approx k_i$. Thus the analysis framework in the Supplementary Note 1 and the branch efficiency analysis (Supplementary Note 2.2) are valid for this case.

### Supplementary Note 6.2 Cooperative effect

Chemical reactions are subject to multiple regulations in the metabolite network. Allosteric enzymes, for instance, can introduce Hill coefficients in the kinetics of chemical reactions[5]. Consider the following scheme:

$$nS_i + E_i \underset{d_i}{\overset{a_i}{\rightleftharpoons}} E_i \cdot nS_i \overset{k_i^{cat}}{\longrightarrow} E_i + nS_{i+1}.$$

The reaction rate $v_i$ follows (assuming that $[S_i] \gg [E_i \cdot nS_i]$):

$$v_i = k_i^{cat} \frac{[S_i]^n}{[S_i]^n + K_i}[E_i], \quad (28)$$

where $K_i \equiv (d_i+k_i)/a_i$. Then, the substrate quality of $S_i$: $\kappa_i = k_i \dfrac{[S_i]^n}{[S_i]^n + K_i}$ (see Supplementary Equation 6). When $\varepsilon$ is maximized, $[S_i] > K_M^{-n}$ and $\kappa_i \approx k_i^{\rightarrow}$, which means that cooperative effect applies to analysis framework in Supplementary Note 1 and the branch efficiency analysis (Supplementary Note 2.2).

**Supplementary Note 6.3 Enzyme inhibitions by metabolites**

There are two general classes of enzyme inhibitors: reversible inhibitors and irreversible inhibitors.

Irreversible inhibitors are mostly small molecules. For metabolites, in principle, are allowed to take this role by inhibiting metabolic reactions of other branches (e.g. catabolic repression in the case of diauxie); however, as we show it below, they are quite unlikely to take this role in their own metabolic branches. Supposing that $S_j$ is a downstream metabolite of reaction $S_i \rightarrow S_{i+1}$ (with $E_i$ is the catalyzing enzyme), while functioning as an irreversible inhibitor:

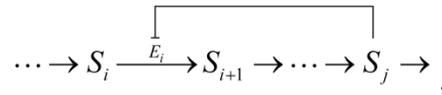

where $S_j$ inhibit $E_i$ irreversibly with reaction:

$$S_j + E_i \rightarrow S_j \cdot E_i.$$

Here the enzyme within $S_j \cdot E_i$ is inactive. Note that $S_j$ is accumulated as long as there is active form of $E_i$, eventually $[S_j] \gg [E_i]$. This mechanism would shut down the metabolic flow through this branch and thus is unlikely to exist.

Reversible inhibitions, generically, have three types[5]: (a) Competitive inhibition

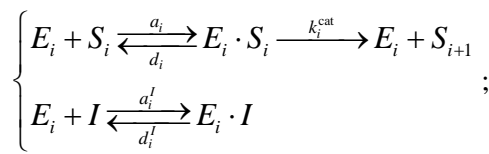

where $I$ denotes the enzyme inhibitor; (b) Uncompetitive inhibition

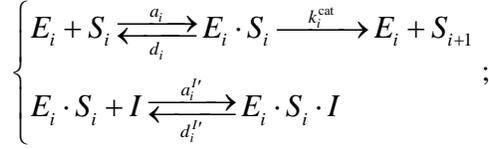

and (c) Mixed inhibition

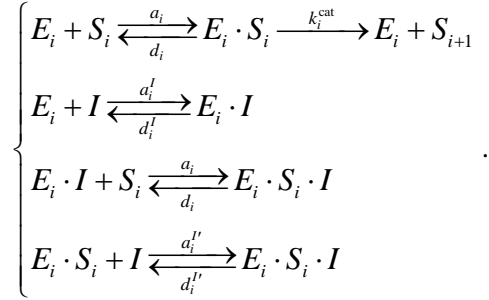

Here, competitive inhibition is common in metabolic network (e.g. ATP regulates phosphofructokinase-1 in glycolysis) while the two other types have not been observed for enzymes with a single substrates[5]. For competitive inhibition, the Michaelis-Menten kinetics is reshaped into[5]

$$v_i = k_i^{cat} \frac{[S_i]}{[S_i] + \gamma K_i}[E_i],  \quad (29)$$

where $K_i \equiv (d_i + k_i^{cat})/a_i$, $\gamma \equiv 1 + \frac{[I]}{K_I}$ and $K_I \equiv d_i^I/a_i^I$. $K_I$ should be not very small since the inhibition is reversible. Then, the substrate quality of $S_i$: $\kappa_i = k_i \frac{[S_i]}{[S_i] + \gamma K_i}$ (see Supplementary Equation 6). Supposing that metabolite $S_j$ is the inhibitor $I$, when $\varepsilon$ is maximized, $[S_i] \gg \gamma K_i = K_i + \frac{K_i}{K_I}[S_j]$ while $[S_j] \gg K_j$, then $\kappa_i \approx k_i$ and $\kappa_j \approx k_j$, which means that analysis framework in Supplementary Note 1 and the branch efficiency analysis (Supplementary Note 2.2) are still applicable.

For uncompetitive inhibition, the Michaelis-Menten kinetics becomes

$$v_i = k_i^{cat} \frac{[S_i]}{\gamma'[S_i] + K_i}[E_i],  \quad (30)$$

where $\gamma' \equiv 1 + \frac{[I]}{K'_I}$, $K'_I \equiv d''_i / a''_i$. In the case of mixed inhibition, the rate equation is

$$v_i = k_i^{cat} \frac{[S_i]}{\gamma'[S_i] + \gamma K_i}[E_i]. \tag{31}$$

In these two cases (reversible inhibition type b &c), however, $\kappa_i$ is clearly dependent on the concentration of $I$. We discuss more general regulations below including these two cases.

**Supplementary Note 6.4 Enzyme regulations by metabolites that permits any function form**

Here we assume that reaction rates $v_i$ depends linearly on $[E_i]$, all forms of metabolic regulations are permitted that $\kappa_i$ (the substrate quality of $S_i$) can be influenced by any metabolites: i.e. $\kappa_i = \kappa_i([\mathbf{S}])$, where $[\mathbf{S}] = ([S_1], [S_2], ..., [S_N])$. This case has been studied with the optimal condition corresponds to elementary flux mode[2, 23, 24]. For our purpose here, when $\varepsilon$ is maximized, the approximation of $\kappa_i \approx k_i$ is no longer ensured. However, every metabolite owns an optimal case specific concentration: $[\mathbf{S}] = [\mathbf{S}^0]$ (i.e. $([S_1], [S_2], ..., [S_N]) = ([S_1^0], [S_2^0], ..., [S_N^0])$. $[\mathbf{S}^0]$ is culturing medium dependent, yet unique for a given medium with a given nutrient concentration, and thus $\kappa_i = \kappa_i([\mathbf{S}^0])$. In this case, we can regard $\kappa_i([\mathbf{S}^0])$ as a medium (with a given nutrient concentration) specific parameter at optimal conditions, branch efficiency analysis qualitatively applies (e.g In Fig.2b and Supplementary Note 2.2, there are fixed value for $\varepsilon_{X \to M}$ and $\varepsilon_{Y \to M}$, yet unable to obtain specifically).

Using optimization principle (see Supplementary Note 1.1) combined with topology features of metabolic network, we can obtain the following qualitative behavior (agree with elementary flux mode[2, 23]): in the case of Supplementary Fig. 1b, either *A*1 or *A*2 will be utilized depending on the growth rate of individual mediums, yet unable to predict the turning point (or ratio sensing behavior); in the case of Supplementary Fig. 1d, three strategies (using only *A*; using only *B*; or using *A* and *B*) are permitted, yet unable to predict if *A* and *B* would be co-utilized, neither does the carbon supply percentage in cases of co-utilization.

**Supplementary Note 6.5 Exceptional cases**

On qualitative aspect, there are two exceptional cases of the branch efficiency analysis (Supplementary Note 2.2) and optimal condition no longer corresponds to elementary flux mode: (a) Overlapping enzymes; (b) Enzyme regulations among each other.

Case (a):

Consider the following scheme:

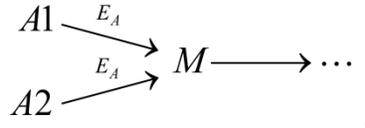

where $E_A$, a single enzyme (or transporter) catalyzing two distinct reactions. Although the topology here is similar to that of Supplementary Fig. 1b, $A1$ and $A2$ would be co-utilized when $\varepsilon$ is maximized.

Case (b):

Conceptually, activity of an enzyme is possible to be influenced by other enzymes. Consider the following scheme:

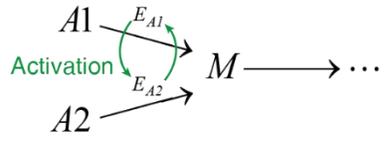

$E_{A1}$ and $E_{A2}$ are carrier enzymes of $A1$ and $A2$, respectively. Specifically:

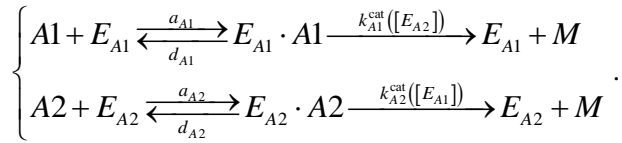

We consider the following imaginary form of $k_{Ai}^{\text{cat}}$ ($i$=1, 2): $k_{A1}^{\text{cat}}([E_{A2}]) = k_{A1}^{\text{L}} + k_{A1}^{\text{H}} \cdot [E_{A2}]$, $k_{A2}^{\text{cat}}([E_{A1}]) = k_{A2}^{\text{L}} + k_{A2}^{\text{H}}[E_{A1}]$ ($k_{Ai}^{\text{L}}, k_{Ai}^{\text{H}} > 0$ and assume $k_{Ai}^{\text{H}} \gg k_{Ai}^{\text{L}}$), where enzymes can promote the activities of each other. The flux from $A1$ to $M$ is $v_{A1} = k_{A1}^{\text{cat}}(E_{A2}) \frac{[A1][E_{A1}]}{[A1] + K_{A1}}$ while the flux from A2 to M is $v_{A2} = k_{A2}^{cat}(E_{A1}) \frac{[A2][E_{A2}]}{[A2]+K_{A2}}$ (we assume Michaelis–Menten kinetics).

Denote $\frac{[Ai]}{[Ai]+K_{Ai}}$ as $f_{[Ai]}$, which is independent of $E_{Ai}$. Then $v_{A1} = k_{A1}^{cat}(E_{A2}) \cdot [E_{A1}] \cdot f_{[A1]}$, and $v_{A2} = k_{A2}^{cat}(E_{A1}) \cdot [E_{A2}] \cdot f_{[A2]}$ Here, $\Phi_{tot} = V_{cell} \cdot ([E_{A1}] \cdot n_{E_{A1}} + [E_{A2}] \cdot n_{E_{A2}})$, while $J_{tot} \equiv V_{cell} \cdot (v_{A1} + v_{A2})$. According to Supplementary Equation 1,

$$\varepsilon = \frac{k_{A1}^L \cdot [E_{A1}] \cdot f_{[A1]} + k_{A2}^L \cdot [E_{A2}] \cdot f_{[A2]} + (k_{A1}^H \cdot f_{[A1]} + k_{A2}^H \cdot f_{[A2]}) \cdot [E_{A1}] \cdot [E_{A2}]}{[E_{A1}] \cdot n_{E_{A1}} + [E_{A2}] \cdot n_{E_{A2}}}, \qquad (32)$$

when $k_{Ai}^H \gg k_{Ai}^L$, at the peaking point of $\varepsilon$, $[E_{A1}], [E_{A2}] > 0$. In this scenario, the metabolic topology is similar to Supplementary Fig. 1b, yet A1 and A2 can be co-utilized when $\varepsilon$ is maximized.

## Supplementary Note 7. Summary and discussions on the application scope of our analysis framework

Our analysis framework (Supplementary Note 1, and thus the branch efficiency analysis in Supplementary Note 2.2) is based on irreversible reactions: when $\varepsilon$ is maximized, $\kappa_i \approx k_i$ for intermediate metabolites (observed for most metabolite in *E. coli*[9, 10]), with the knowledge of specific activity (defined as enzyme turnover number divided by molecular weight) of catabolic enzymes, we obtain substrate quality $\kappa_i$ and can predict the growth behavior of microbes on mixed carbon sources. With this framework, we quantitatively explain the phenomenon of diauxie versus co-utilization; predictions of carbon supply percentage in various combinations of mixtures agree well with experimental results.

In Supplementary Note 5, we show that this framework applies to reversible reactions. In Supplementary Note 6, we demonstrate that this framework is broadly applicable to cases such as reversible reactions, enzyme concentration dependent reaction rate, allosteric enzymes, irreversible metabolite inhibitions, and reversible competitive inhibition. When all forms of metabolic regulations are permitted, this framework only qualitatively applies, i.e. qualitatively behavior that is possible to show up based on the metabolic topology (Supplementary Note 6.4). However, conceptually or in practice, there are two exceptional cases for this framework:

Bi-substrates transporters/enzymes (e.g. glucose transporters in *E. coli* can co-transport mannose[25]), and enzymes regulation among each other (Supplementary Note 6.5). Nevertheless, we can consider these effects specifically when they are involves.

Overall, our analysis framework (Supplementary Note 1) can be broadly applicable for microbial studies and useful in quantitatively explaining why and how microbes make the choices when facing multiple carbon sources.

**Supplementary Figures**

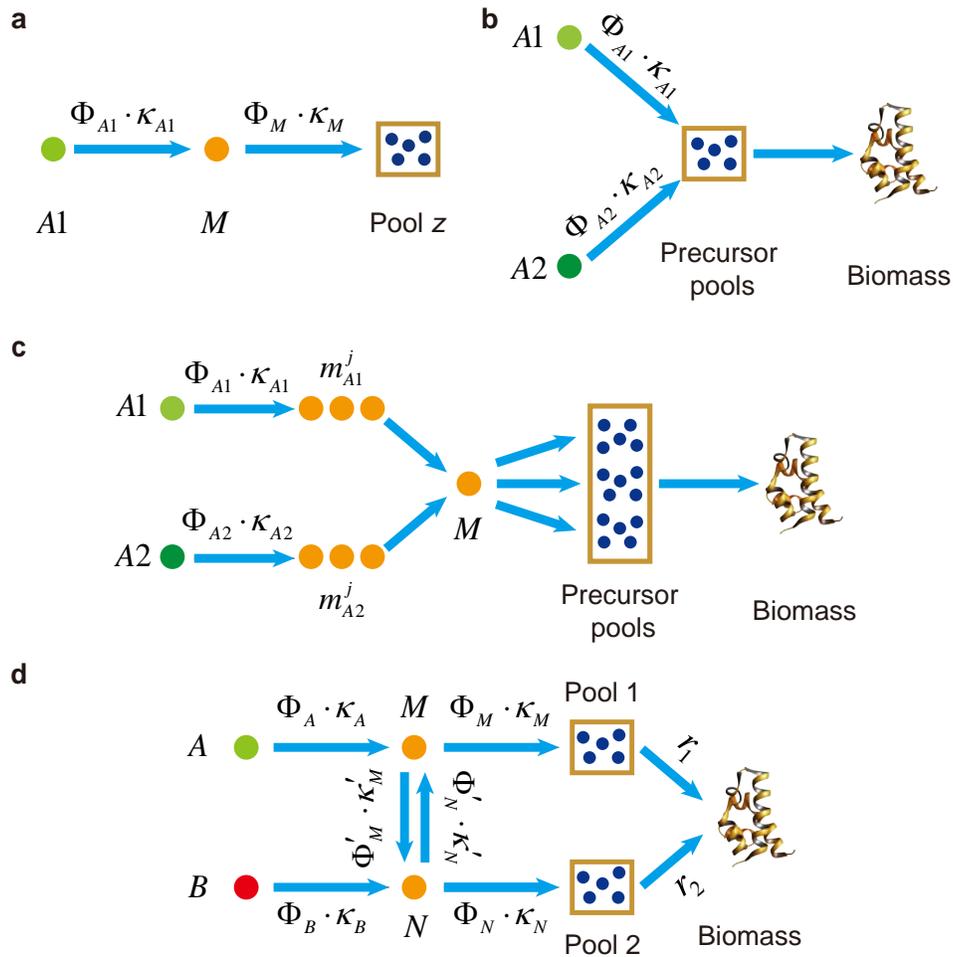

**Supplementary Figure 1. Coarse-grained models of metabolism and biomass production.**

(**a**) A coarse-grained metabolic model with one intermediate node.

(**b**) Minimal model of diauxie. The carbon sources $A1$ or $A2$ or both can supply the precursor pools. The cell grows faster if only the more efficient source is utilized.

(**c**) Topology of metabolic network with two Group A sources. The two carbon flux pathways from sources $A1$ and $A2$ can have multiple intermediate nodes (metabolites) $m_{A1}^j$ and $m_{A2}^j$ before merging to a common node $M$, after which the flux is diverted to various precursor pools.

(**d**) Minimal model of co-utilization. In synthesizing biomass, the two precursor pools supply $r_1$ and $r_2$ carbon flux, respectively. Either pool can draw flux from either of the two sources A and B. Under certain conditions, it is optimal for different sources to supply different pools.

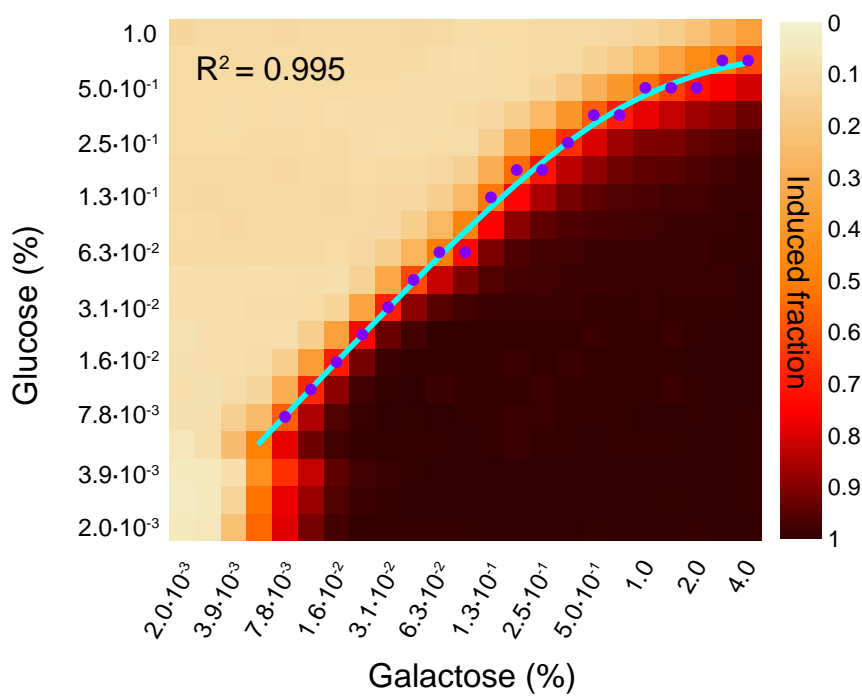

**Supplementary Figure 2. Concentration dependence of turning point.** In the experiment of Escalante-Chong et al.[11], yeast cells were cultured with a mixture of glucose and galactose of various combinations of concentrations. The induction of galactose pathway was measured in single cells with flow cytometry. The heat map represents the fraction of cells with the galactose pathway turned on for given pairs of concentrations (reproduced with permission). The purple dots indicate the glucose concentration at which the induction fraction is at or just above 0.5 for given galactose concentration. The solid line is a fit with Supplementary Equation 14 ($R^2 = 0.995$, $\delta = 0.8256$ and $\Delta = 0.8052$).

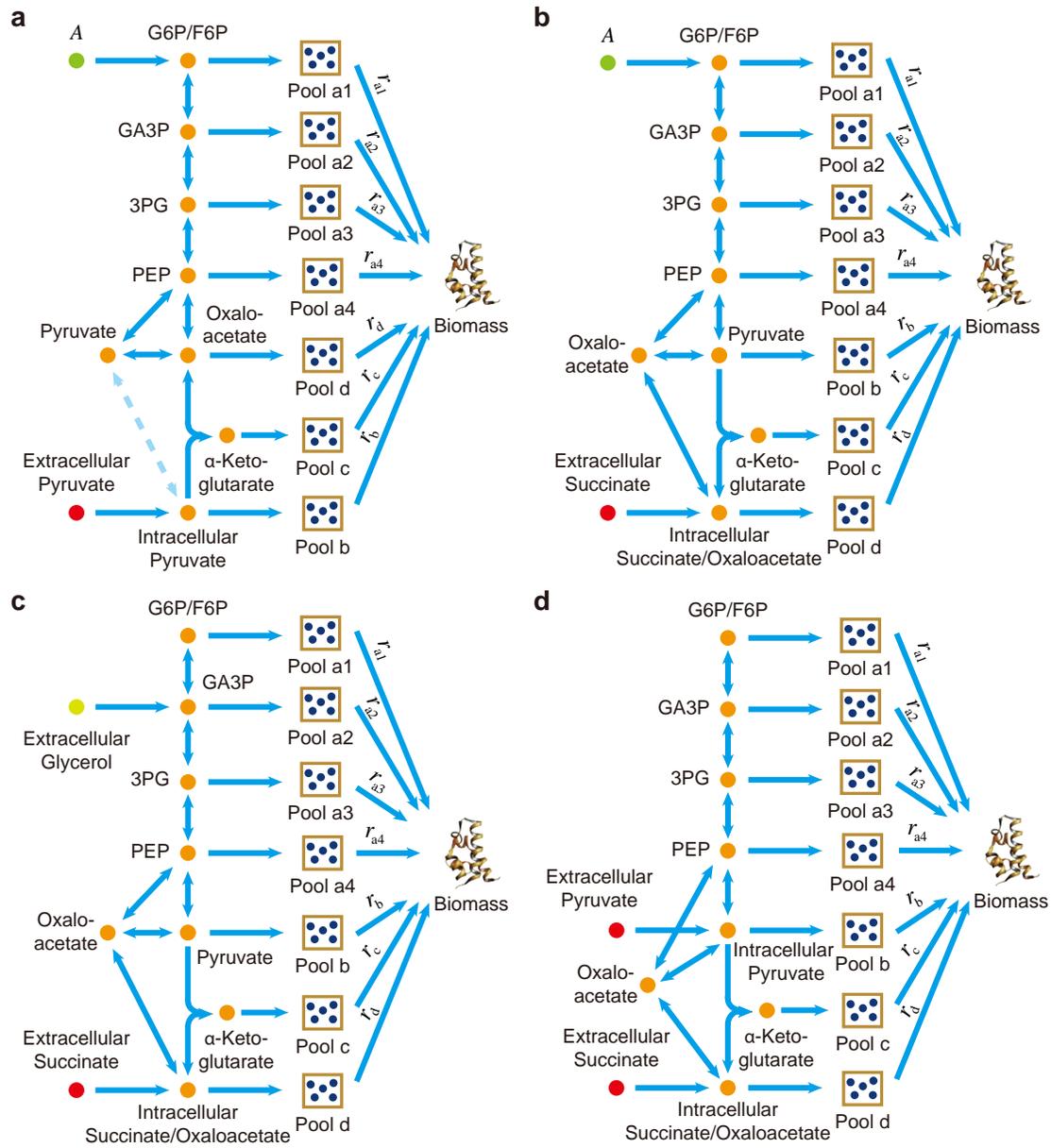

**Supplementary Figure 3. Topology of the metabolic network when a Group B source is mixed with a Group A source or with another Group B source.**

(**a**) Pyruvate (a Group B source) is mixed with a Group A source.

(**b**) Succinate (a Group B source) is mixed with a Group A source.

(**c**) Succinate (a Group B source) is mixed with Glycerol (a Group A source).

(**d**) Pyruvate (a Group B source) is mixed with Succinate (a Group B source).

See Supplementary Note 1.3 for the classifications of Pools a1-a4, b-d.

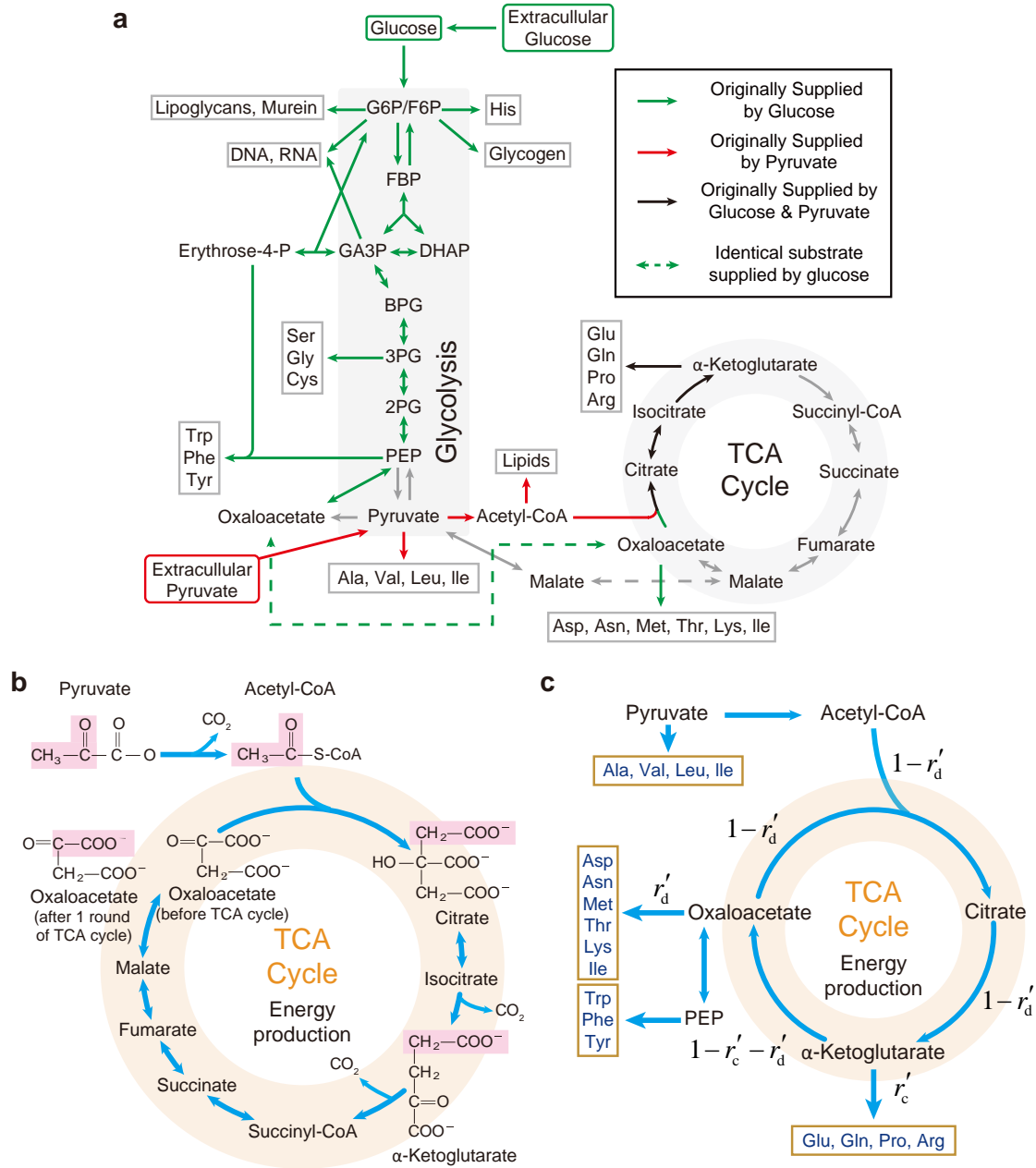

**Supplementary Figure 4. Pool suppliers in the case of co-utilization.** (**a**) Original pool suppliers (Model predictions, see Supplementary Table3) in the case of glucose-pyruvate (both with saturated concentrations) co-utilization. Metabolites connected by green arrows are originally supplied by glucose, those connected by red arrows are originally supplied by pyruvate, and those connected by black arrows are originally supplied by both glucose and pyruvate. (**b**) & (**c**) Influence of the TCA cycle on the pool suppliers in practice. (**b**) Reactions of the TCA cycle. Through a round of TCA cycle, half of the carbon atoms of an oxaloacetate molecule are replaced

by that of pyruvate (marked with light pink shades). (**c**) Stoichiometry allocation of the carbon flux in TCA cycle. For a given microbial growth rate, we assume the stoichiometry of carbon flux at oxaloacetate is 1 per unit time $\tau$, $r'_d$ stoichiometry of the flux flows to Pool d (Aspartic acid, etc.), $1-r'_d$ stoichiometry of flux flows to citrate, accompanied with the same stoichiometry of carbon flux joined from Citrate. $r'_c$ stoichiometry of the flux flows to Pool c (Glutamic acid, etc.), with $1-r'_c-r'_d$ stoichiometry of carbon flux flows back to oxaloacetate. To keep sustainable microbial growth, $r'_c+r'_d$ stoichiometry of carbon flux would join from the original supplier of oxaloacetate (entry point of Pool d).

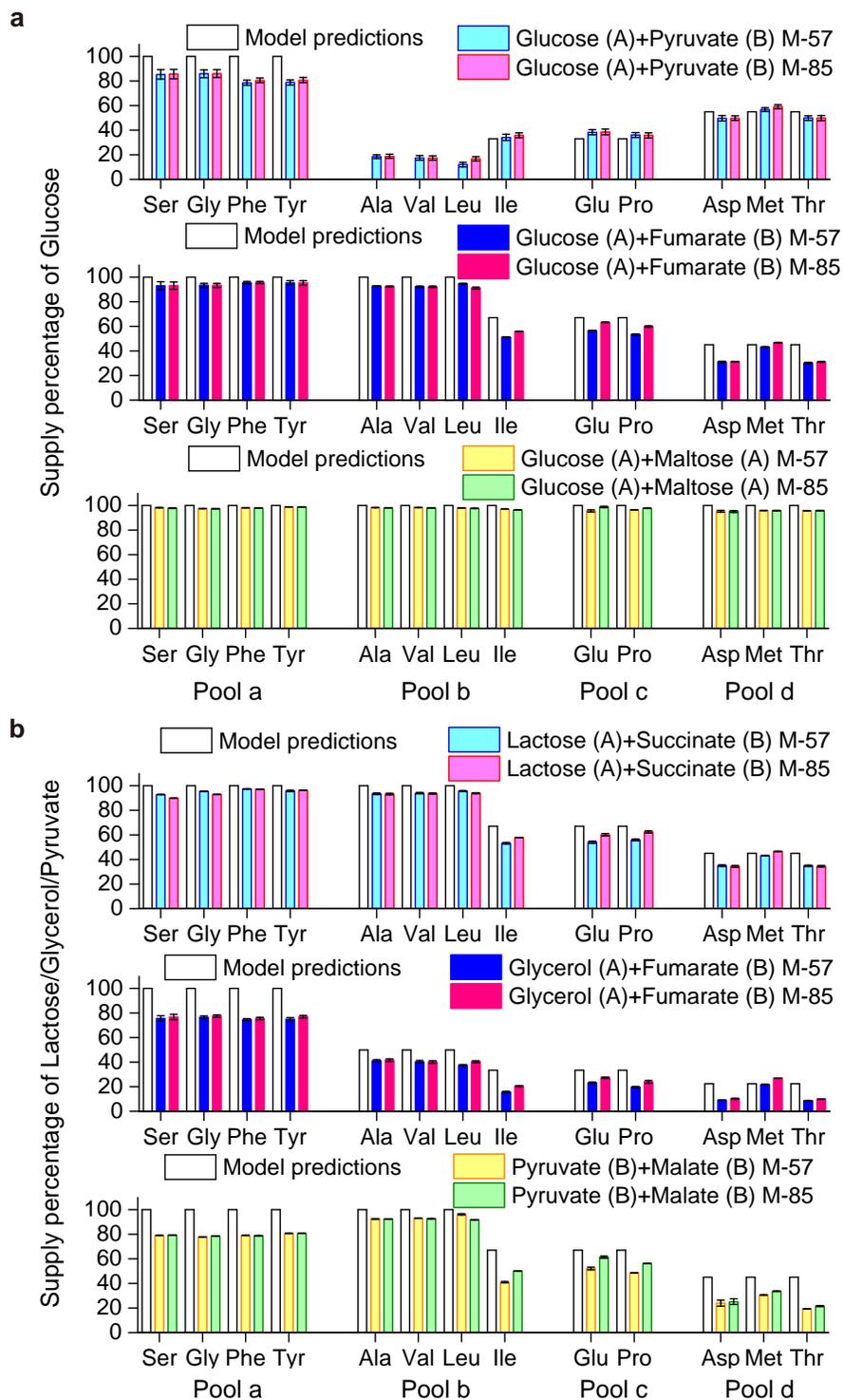

**Supplementary Figure 5. Comparison of pool suppliers determined using experimental data M-57 and M-85.** (a) Cases of glucose mixed with another carbon source. (b) Cases of lactose, glycerol or pyruvate mixed with a Group B carbon source. Leu M-15 and Ile M-15 data are used in the M-57 results (see Methods for details). In all cases, there is no significant difference

between results obtained using data M-57 and data M-85. Model predictions are marked with hollow bars while experimental results are marked with color bars. Error bars represent standard deviations. Source data are provided as a Source Data file.

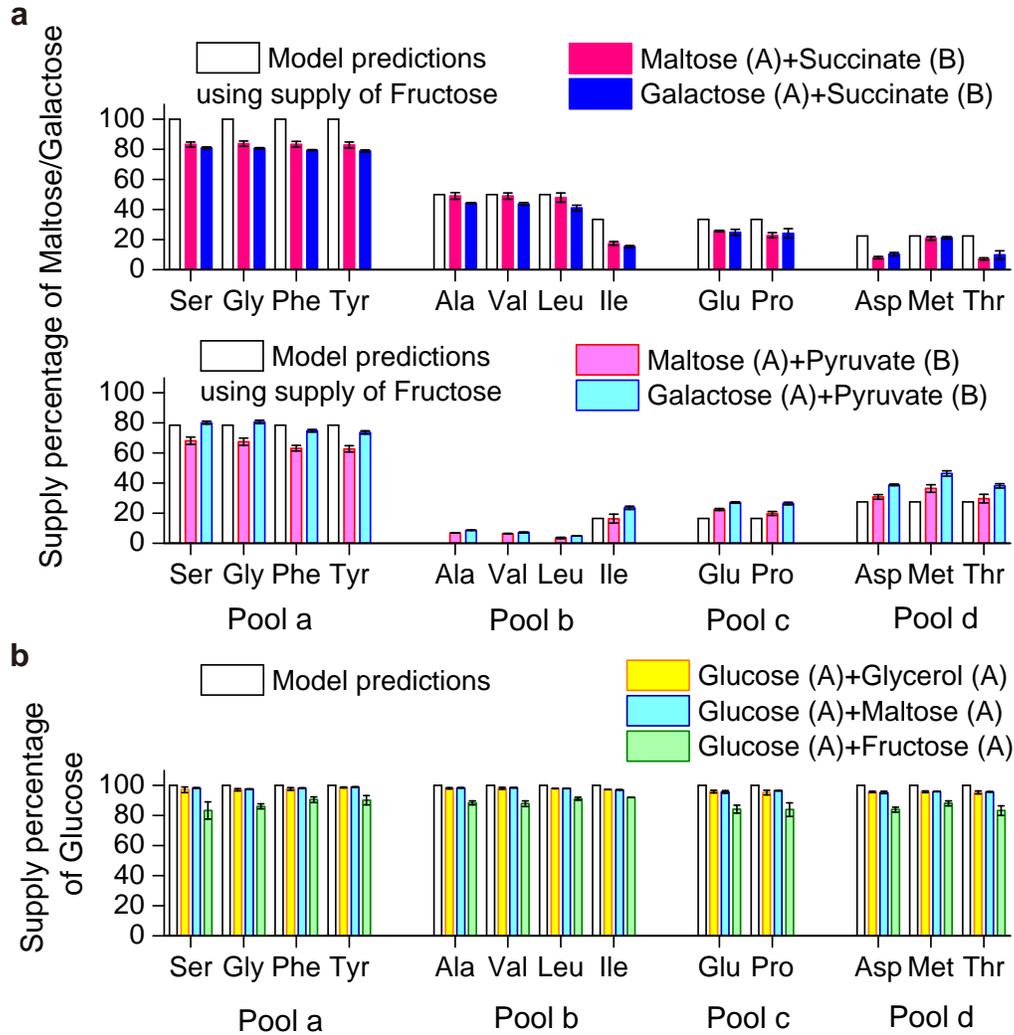

**Supplementary Figure 6. Suppliers of precursor pools on mixed carbon sources. (a)** Cases of co-utilization (A+B). Here the supply percentage of maltose and galactose in each case are compared with model predictions using supply of fructose (Fructose+Succinate or Fructose+Pyruvate), since some biochemical parameters of the catabolic enzymes of maltose, and galactose are unable to find, while the topology and supply percentage are quite similar to that of fructose. **(b)** Cases of diauxie (A+A): glucose mixed with glycerol, maltose, and fructose. For glucose mixed with glycerol and maltose (non-PTS sugars), all pools are supplied by glucose, which are perfect cases of diauxie. For glucose mixed with fructose (a PTS sugar), the majority of carbon (>83%) in all pools are supplied by glucose, with small portions supplied by fructose, which might due to imperfect molecular inhibition. Model predictions are marked with hollow bars while experimental results are marked with color bars. Error bars represent standard deviations. Source data are provided as a Source Data file.

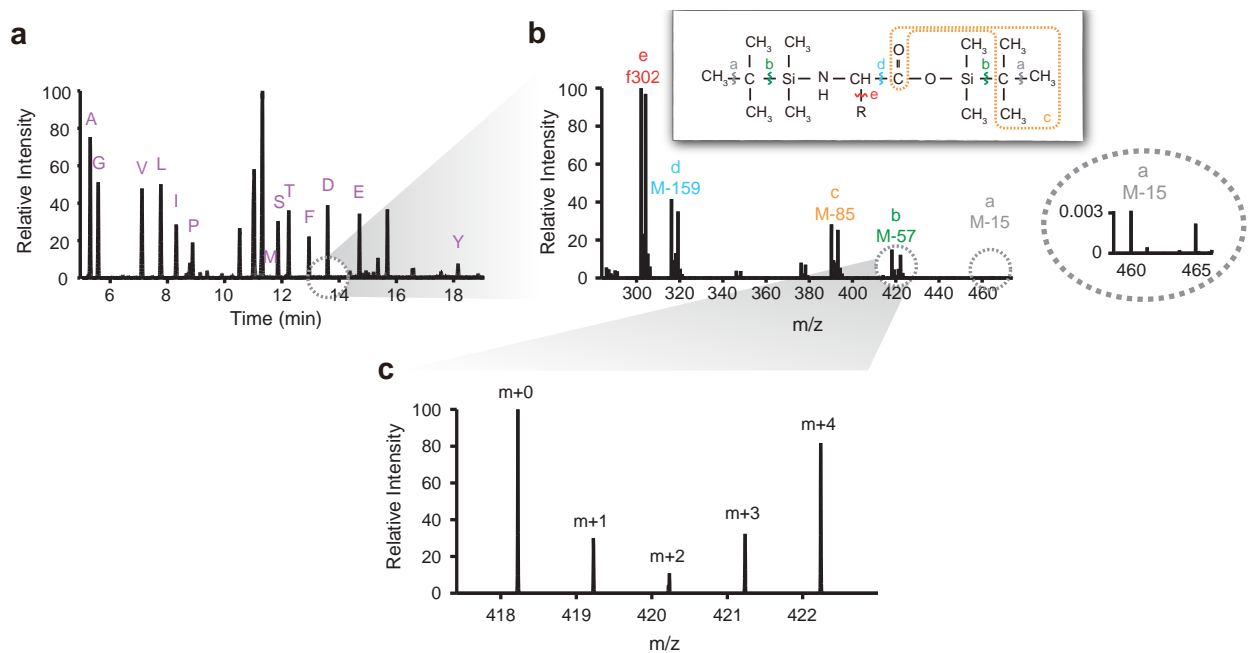

**Supplementary Figure 7. Measurement of $^{13}$C labeling percentage by GC-MS analysis[26, 27].**
(**a**) The gas chromatogram of derivatized amino acid from *E.coli* hydrolysate, annotated by their one-letter abbreviation. These different amino acids were separated by GC according to their different retention time. Inset of (**b**) The structure of derivatized amino acids. During ionization, the derivatized amino acids were fragmented, either cracking at the wave line or losing the atoms in the box, resulting in various fragments, including a: M-15, b: M-57, c: M-85, d: M-159 and e: f302. These different fragments of the same amino acid were further separated and analyzed by MS. (**b**) Integrated mass spectrum over the full GC peak of derivatized aspartic acid (retention time from 13.57~13.63 min). There are five different fragments of derivatized aspartic acid detected by the MS. For every fragment, several mass isotopomer peaks are detected, referring to the same chemical structure but incorporating different number of $^{13}$C atoms. (**c**). The mass spectrum of M-57 fragment of derivatized aspartic acid. There are five mass isotopomer peaks of M-57 fragment. m + 0 denotes the mass isotopomer that contains no $^{13}$C atoms. m + 1, m + 2, m + 3 and m + 4 denote the mass isotopomer that contains 1 to 4 $^{13}$C atoms respectively.

## Supplementary Tables

**Supplementary Table 1.** $k_{cat}$ and molecular weight (MW) reference data of *E. coli*.

| Reaction | Enzyme/ Transporter | $k_{cat}$ (s$^{-1}$) | MW (kDa) | References |
|---|---|---|---|---|
| Glucose → Glucose-6P | Glucokinase | $4.1 \times 10^2$ | $7.0 \times 10$ | 28-30 |
| Glucose-6P ↔ Fructose-6P | Glucose-6-phosphate isomerase | $2.6 \times 10^2$ | $1.2 \times 10^2$ | 31, 32 |
| Fructose-6P → Fructose-1,6P | Phosphofructokinase | $4.4 \times 10^2$ | $1.4 \times 10^2$ | 33, 34 |
| Fructose-1,6P → Fructose-6P | Fructose-1,6-bisphosphatase | $2.0 \times 10$ | $3.6 \times 10$ | 35, 36 |
| Fructose-1,6P ↔ Glyceraldehyde 3-phosphate + Dihydroxyacetone phosphate | Fructose-bisphosphate aldolase | $1.4 \times 10$ | $7.8 \times 10$ | 37, 38 |
| Dihydroxyacetone phosphate ↔ Glyceraldehyde 3-phosphate | Triosephosphate Isomerase | $4.3 \times 10^2$ | $5.4 \times 10$ | 39, 40 |
| Glyceraldehyde 3-phosphate ↔ 1,3-Bisphosphoglycerate | Glyceraldehyde-3-ph-osphate dehydrogenase | $9.5 \times 10$ | $1.4 \times 10^2$ | 41, 42 |
| 1,3-Bisphosphoglycerate ↔ 3-Phosphoglycerate | Phosphoglycerate kinase | $3.5 \times 10^2$ | $4.4 \times 10$ | 43, 44 |
| 3-Phosphoglycerate ↔ 2-Phosphoglycerate | Phosphoglycerate mutase | $3.3 \times 10^2$ | $4.9 \times 10$ | 45 |
| 2-Phosphoglycerate ↔ Phosphoenolpyruvate | Enolase | $2.2 \times 10^2$ | $9.0 \times 10$ | 46, 47 |
| Phosphoenolpyruvate → Pyruvate | Pyruvate kinase | $5.0 \times 10^2$ | $2.4 \times 10^2$ | 48 |
| Pyruvate → Acetyl-CoA | Pyruvate dehydrogenase | $1.2 \times 10^2$ | $1.0 \times 10^2$ | 49 |
| Oxaloacetate+Acetyl-CoA → Citrate | Citrate synthase | $2.4 \times 10^2$ | $9.7 \times 10$ | 50, 51 |
| Citrate ↔ Isocitrate | Aconitate hydratase | $7.0 \times 10$ | $9.4 \times 10$ | 52, 53 |
| Isocitrate → α-Ketoglutarate | Isocitrate dehydrogenase | $2.0 \times 10^2$ | $9.5 \times 10$ | 42, 54, 55 |
| α-Ketoglutarate → Succinyl-CoA | α-Ketoglutarate dehydrogenase complex E1 component | $1.5 \times 10^2$ | $1.9 \times 10^2$ | 56, 57 |
| Succinyl-CoA ↔ Succinate | Succinyl-CoA synthetase | $9.1 \times 10$ | $1.6 \times 10^2$ | 58 |
| Succinate → Fumarate | Succinate dehydrogenase | $1.1 \times 10^2$ | $1.0 \times 10^2$ | 59, 60 |
| Fumarate → Succinate | Fumarate reductase | $2.5 \times 10^2$ | $9.3 \times 10$ | 59, 61 |
| Fumarate ↔ Malate | Fumarase | $1.2 \times 10^3$ | $2.0 \times 10^2$ | 62, 63 |
| Malate ↔ Oxaloacetate | Malate dehydrogenase | $5.5 \times 10^2$ | $6.1 \times 10$ | 64 |
| Phosphoenolpyruvate → Oxaloacetate | Phosphoenolpyruvate carboxylase | $1.5 \times 10^2$ | $4.0 \times 10^2$ | 65, 66 |
| Oxaloacetate → Phosphoenolpyruvate | Phosphoenolpyruvate carboxykinase | 4.3 | $6.0 \times 10$ | 67-69 |
| Malate → Pyruvate | Malic enzyme | $8.3 \times 10$ | $2.7 \times 10^2$ | 70, 71 |
| Pyruvate → Malate | Malic enzyme | 2.9 | $2.7 \times 10^2$ | 70, 71 |
| Pyruvate → Oxaloacetate | - | - | - | - |
| Pyruvate → Phosphoenolpyruvate | Pyruvate, water dikinase | $3.5 \times 10$ | $2.5 \times 10^2$ | 72 |
| Extracellular Glucose → Glucose-6P | Glucose-specific PTS enzyme | $1 \times 10^2$ | $5.0 \times 10$ | 12, 73-76 |
| Lactose membrane transport | Lactose permease | $5 \times 10$ | $4.6 \times 10$ | 77, 78 |
| Lactose→Glucose+Galactose | β-galactosidase | $6.4 \times 10^2$ | $4.6 \times 10^2$ | Estimated 79, 80 |
| Extracellular Fructose →Fructose-6P† | Fructose-specific PTS enzyme | 17 | $5.8 \times 10$ | Estimated 12, 76 |
| Glycerol membrane transport | Glycerol facilitator | 4 | $2.5 \times 10$ | Estimated 12, 81-83 |
| Glycerol →Glycerol-3-phosphate | Glycerol kinase | $1.4 \times 10^2$ | $2.1 \times 10^2$ | 84, 85 |
| Glycerol-3-phosphate ↔ Dihydroxyacetone phosphate | Glycerol-3-phosphate dehydrogenase | $6.8 \times 10$ | $1.1 \times 10^2$ | 86 |
| Pyruvate membrane transport | Pyruvate transporter | 9 | $4 \times 10$ | Estimated 12, 42, 87-89 |
| Oxaloacetate membrane transport | Oxaloacetate transporter | 5 | $4 \times 10$ | Estimated 12, 42, 90-92 |
| Succinate membrane transport | Succinate transporter | $5.1^*$ | $4 \times 10$ | Estimated 12, 42, 92-94 |
| Fumarate membrane transport | Fumarate transporter | $4.6^*$ | $4 \times 10$ | Estimated 12, 42, 92, 95 |
| Malate membrane transport | Malate transporter | $4.5^*$ | $4 \times 10$ | Estimated 12, 42, 92, 95 |

† The uptake of extracellular fructose is different (Extracellular Fructose→Fructose-1P) at low concentration (<2mM).

* Estimated value around 5 with fitting decimal digits since that the growth rate of *E.coli* are similar in culturing medium with succinate, fumarate or malate as the single carbon source (see Supplementary Table 5).

**Supplementary Table 2. Substrate branch efficiency of Pools a1-a4, Pool b and Pool d of *E. coli*.** (With enzyme MW normalization unit 100 kDa, branch efficiency unit $s^{-1}$, and carbon sources of saturated concentrations.)

| Substrate (Sub) | Pool a1 | Pool a2 | Pool a3 | Pool a4 | Pool b | Pool d |
|---|---|---|---|---|---|---|
| | Ser, Gly, Cys, Trp, Phe, Tyr, His; precursors of RNA, DNA, Glycogen, Lipoglycans, Murein. | | | | Ala, Val, Leu, Ile†; precursors of Lipids. | Asp, Asn, Met, Thr, Lys, Ile‡ |
| | $\varepsilon_{Sub \to F6P}$ | $\varepsilon_{Sub \to GA3P}$ | $\varepsilon_{Sub \to 3PG}$ | $\varepsilon_{Sub \to PEP}$ | $\varepsilon_{Sub \to pyruvate}$ | $\varepsilon_{Sub \to oxaloacetate}$ |
| Glucose | 104 | 14.6 | 11.7 | 10.9 | 10.4 | 8.5 |
| Lactose | 44 | 12.3 | 10.1 | 9.6 | 9.2 | 7.6 |
| Fructose | 29 | 10.8 | 9.1 | 8.6 | 8.3 | 7.0 |
| Glycerol | 5.9 | 10.5 | 9.0 | 8.6 | 8.2 | 7.0 |
| Pyruvate | 4.7 | 7.3 | 8.2 | 8.6 | 22.5 | 7.0 |
| *Oxaloacetate | 3.6 | 4.8 | 5.2 | 5.4 | 8.8 | 12.5 |
| Malate | 3.5 | 4.7 | 5.0 | 5.2 | 8.2 | 11.1 |
| Fumarate | 3.5 | 4.7 | 5.0 | 5.2 | 8.2 | 11.1 |
| Succinate | 3.5 | 4.7 | 5.0 | 5.2 | 8.2 | 11.1 |

*Oxaloacetate can quickly decompose to pyruvate and $CO_2$ spontaneously[16] in solution (Supplementary Note 4.7), which causes the pool suppliers in cases involving oxaloacetate different from that predicted using the branch efficiency.

†Isoleucine is supplied by joint fluxes from pyruvate and oxaloacetate (Supplementary Note 4.2).

**Supplementary Table 3. Predicted original pool suppliers of *E. coli* in the cases of co-utilization.** The original suppliers of Pools a, b, d are determined by the branch efficiency listed in Supplementary Table 2. Owing to the Influence of TCA cycle, the pool supplier in practice can be different from its original supplier.

| No. | Mixed Substrates | Pools a1, a2, a3, a4<br>Ser, Gly, Cys, Trp, Phe, Tyr, His; precursors of RNA, DNA, Glycogen, Lipoglycans, Murein.<br>**Original Supplier** | Pool b<br>Ala, Val, Leu, Ile; precursors of Lipids<br>**Original Supplier** | Pool d<br>Asp, Asn, Met, Thr, Lys, Ile<br>**Original Supplier** |
|---|---|---|---|---|
| 1 | Glucose/ Lactose + Pyruvate | Glucose/ Lactose | Pyruvate | Glucose/ Lactose |
| 2 | Glucose/ Lactose + Succinate/ Malate/ Fumarate | Glucose/ Lactose | Glucose/ Lactose | Succinate/ Malate/ Fumarate |
| 3 | Fructose/Glycerol + Pyruvate | Fructose/Glycerol (a1, a2, a3, 50%* a4), Pyruvate (50%* a4) | Pyruvate | Fructose/Glycerol (50%*), Pyruvate (50%*) |
| 4 | Fructose/Glycerol +Succinate/ Malate/ Fumarate | Fructose/Glycerol | Fructose/Glycerol (50%*), Succinate/ Malate/ Fumarate (50%*) | Succinate/ Malate/ Fumarate |
| 5 | Pyruvate +Succinate/ Malate/ Fumarate | Pyruvate | Pyruvate | Succinate/ Malate/ Fumarate |
| 6 | Succinate +Malate† | Succinate(50%*), Malate(50%*) | Succinate(50%*), Malate(50%*) | Succinate(50%*), Malate(50%*) |

*Roughly identical branch efficiencies from both substrates.

†Succinate and malate own shared membrane transporters.

**Supplementary Table 4. Pool suppliers of *E. coli* predicted to be observed in the cases of co-utilization under aerobic conditions.** The pool suppliers in practice are influence by joint fluxes as well as the TCA cycle (see Supplementary Notes 4.1-4.5 for details).

| No. | Mixed Substrates | Pools a1, a2, a3, a4 | Pool b | Pool c | Pool d |
|---|---|---|---|---|---|
| | | Ser$^\dagger$, Gly$^\dagger$, Cys$^\dagger$, Trp, Phe, Tyr, His; precursors of RNA, DNA, Glycogen, Lipoglycans, Murein. | Ala, Val, Leu, Ile$^*$; precursors of Lipids | Glu, Gln, Pro, Arg | Asp, Asn, Met, Thr, Lys, Ile$^*$ |
| | | Suppliers | Suppliers | Suppliers | Suppliers |
| 1 | Glucose/ Lactose + Pyruvate | Glucose/ Lactose | Pyruvate | Glucose/ Lactose (33%), Pyruvate (67%) | Glucose/ Lactose (55%), Pyruvate (45%) |
| 2 | Glucose/ Lactose + Succinate/ Malate/ Fumarate | Glucose/ Lactose | Glucose/ Lactose | Glucose/ Lactose (67%), Succinate/ Malate/ Fumarate (33%) | Glucose/ Lactose (45%), Succinate/ Malate/ Fumarate (55%) |
| 3 | Fructose/Glycerol + Pyruvate | Fructose/Glycerol (a1, a2, a3, 50% a4), Pyruvate (50% a4) | Pyruvate | Fructose/Glycerol (16.5%), Pyruvate (83.5%) | Fructose/Glycerol (27.5%), Pyruvate (72.5%) |
| 4 | Fructose/Glycerol +Succinate/ Malate/ Fumarate | Fructose/Glycerol | Fructose/Glycerol (50%), Succinate/ Malate/ Fumarate (50%) | Fructose/Glycerol (33.5%), Succinate/ Malate/ Fumarate (66.5%) | Fructose/Glycerol (22.5%), Succinate/ Malate/ Fumarate (77.5%) |
| 5 | Pyruvate, Succinate/ Malate/ Fumarate | Pyruvate | Pyruvate | Pyruvate (67%), Succinate/ Malate/ Fumarate (33%) | Pyruvate (45%), Succinate/ Malate/ Fumarate (55%) |
| 6 | Succinate, Malate | Succinate(50%), Malate(50%) | Succinate(50%), Malate(50%) | Succinate(50%), Malate(50%) | Succinate(50%), Malate(50%) |

$^\dagger$The supply percentage of serine, glycine and cysteine can be affected by the reversible reactions among the precursors of Pool a2-a4 (Supplementary Note 5.1).

$^*$ 2/5 of carbon atoms in isoleucine in practice supplied by Pools b and 3/5 of carbon atoms in practice supplied by Pools d (Supplementary Note 4.2).

**Supplementary Table 5. Exponential growth rate of *E. coli* NCM3722.**

| Growth rate$^†$(h$^{-1}$) | alone | glucose | lactose | glycerol | fructose | maltose | galactose | pyruvate | succinate |
|---|---|---|---|---|---|---|---|---|---|
| alone | / | 1.51$^{††}$ | 1.54$^{††}$ | 1.04 | 1.11 | 1.18 | 0.83$^*$ | 1.15 | 1.07 |
| pyruvate | 1.15 | 1.38$^{**}$ | 1.58 | 1.45 | 0.99$^{**}$ | 1.21 | 1.40 | / | 1.37 |
| succinate | 1.07 | 1.57 | 1.57 | 1.32 | 1.31 | 1.23 | 1.23 | 1.37 | / |
| fumarate | 1.03 | 1.55 | 1.53 | 1.31 | 1.28 | 1.20 | 1.17 | 1.32 | 0$^‡$ |
| malate | 1.02 | 1.53 | 1.54 | 1.30 | 1.32 | 1.26 | 1.23 | 1.34 | 0.98 |

$^†$Observational error and standard deviation of the growth rate data around or less than 0.07h$^{-1}$. All numbers are averages over three independent experiments. Source data are provided as a Source Data file.

$^*$The growth rate of *E.coli* in galactose alone medium is suboptimal due to *GalS* regulation[96].

$^{**}$Membrane transport of Phosphotransferase System (PTS) sugars (such as glucose and fructose) are coupled with the conversion from PEP to pyruvate[97]. This might lead to suboptimal growth when PTS sugars mixed with pyruvate.

$^{††}$Growth rate of *E.coli* is a bit higher in lactose alone medium than that of glucose (although quite similar, and similar in mixed medium). The existence of *lac* operon might due to that glucose is more wide spread in natural environment.

$^‡$ *E.coli* did not grow in succinate+fumarate and malate+fumarate culture medium in 72 hours.

**Supplementary Table 6. All possible patterns of the original pool suppliers of *E. coli* in mixed carbon sources.** Here we consider all possibilities for the choice of the biochemical parameters ($k_{cat}$ can be any positive values). Predictions based on parameters listed in Supplementary Table 1 are shown in blue text (see Supplementary Table 3 for details).

| Mixed Substrates | Possible Pattern Types | Pools a1, a2, a3, a4 | Pool b | Pool d |
|---|---|---|---|---|
| | | Ser, Gly, Cys, Trp, Phe, Tyr, His; precursors of RNA, DNA, Glycogen, Lipoglycans, Murein. | Ala, Val, Leu, Ile; precursors of Lipids | Asp, Asn, Met, Thr, Lys, Ile |
| | | Original Supplier | Original Supplier | Original Supplier |
| A (Glucose/ Lactose/ Fructose/ Glycerol/ Maltose/ Galactose) + B1 (Pyruvate) | DX* No.1 | A | A | A |
| | DX No.2 | B1 | B1 | B1 |
| | CoU* No.1 | A | B1 | A |
| | CoU No.2 | A | B1 | B1 |
| | CoU No.3 | A (a1, a2, a3), B1 (a4) | B1 | B1 |
| | CoU No.4 | A (a1, a2), B1 (a3, a4) | B1 | B1 |
| | CoU No.5 | A(a1), B1 (a2, a3, a4) | B1 | B1 |
| | CoU No.6 | A | A (50%), B1 (50%) | A |
| | CoU No.7 | A | A (50%), B1 (50%) | A (50%), B1 (50%) |
| | CoU No.8 | A | B1 | A (50%), B1 (50%) |
| | CoU No.9 | A (a1, a2, a3, 50% a4), B1 (50% a4) | B1 | A (50%), B1 (50%) |
| | CoU No.10 | A (a1, a2, a3, 50% a4), B1 (50% a4) | B1 | B1 |
| | CoU No.11 | A (a1, a2, 50% a3), B1 (a4, 50% a3) | B1 | B1 |
| | CoU No.12 | A (a1, 50% a2), B1 (a3, a4, 50% a2) | B1 | B1 |
| | CoU No.13 | A (50% a1), B1 (a2, a3, a4, 50% a1) | B1 | B1 |
| A (Glucose/ Lactose/ Fructose/ Glycerol/ Maltose/ Galactose) +B2 (Succinate/ | DX No.1 | A | A | A |
| | DX No.2 | B2 | B2 | B2 |
| | CoU No.1 | A | A | B2 |
| | CoU No.2 | A | B2 | B2 |
| | CoU No.3 | A | B2 | A |
| | CoU No.4 | A (a1, a2, a3), B2 (a4) | B2 | B2 |
| | CoU No.5 | A (a1, a2), B2 (a3, a4) | B2 | B2 |
| | CoU No.6 | A (a1), B2 (a2, a3, a4) | B2 | B2 |
| Malate/<br>Fumarate) | CoU No.7 | A | A (50%),<br>B2 (50%) | A |
| | CoU No.8 | A | A | A (50%), B2 (50%) |
| | CoU No.9 | A | A (50%), B2 (50%) | B2 |
| | CoU No.10 | A | B2 | A (50%), B2 (50%) |
| | CoU No.11 | A | A (50%), B2 (50%) | A (50%), B2 (50%) |
| | CoU No.12 | A (a1, a2, a3, 50% a4),<br>B2 (50% a4) | B2 | B2 |
| | CoU No.13 | A (a1, a2, a3, 50% a4),<br>B2 (50% a4) | A (50%), B2 (50%) | B2 |
| | CoU No.14 | A (a1, a2, a3, 50% a4),<br>B2 (50% a4) | B2 | A (50%), B2 (50%) |
| | CoU No.15 | A (a1, a2, 50% a3),<br>B2 (a4, 50% a3) | B2 | B2 |
| | CoU No.16 | A (a1, 50% a2),<br>B1 (a3, a4, 50% a2) | B2 | B2 |
| | CoU No.17 | A (a1, 50% a2),<br>B1 (a3, a4, 50% a2) | B2 | B2 |
| B1 (Pyruvate)<br>+B2<br>(Succinate/<br>Malate/<br>Fumarate) | DX No.1 | B1 | B1 | B1 |
| | DX No.2 | B2 | B2 | B2 |
| | CoU No.1 | B1 | B1 | B2 |
| | CoU No.2 | B2 | B1 | B2 |
| | CoU No.3 | B1 | B1 | B1 (50%), B2 (50%) |
| | CoU No.4 | B2 | B1 (50%), B2 (50%) | B2 |
| | CoU No.5 | B1 (50%), B2 (50%) | B1 | B2 |
| | CoU No.6 | B1 (50%), B2 (50%) | B1 (50%), B2 (50%) | B2 |
| | CoU No.7 | B1 (50%), B2 (50%) | B1 | B1 (50%), B2 (50%) |
| | CoU No.8 | B1 (50%), B2 (50%) | B1 (50%), B2 (50%) | B1 (50%), B2 (50%) |
| Succinate +<br>Malate | DX No.1 | Succinate | Succinate | Succinate |
| | DX No.2 | Malate | Malate | Malate |
| | CoU No.1 | Succinate (50%),<br>Malate (50%) | Succinate (50%),<br>Malate (50%) | Succinate (50%),<br>Malate (50%) |

[*]DX denotes diauxie and CoU signifies co-utilization.

**Supplementary Table 7. All possible patterns of the pool suppliers of *E. coli* under aerobic conditions in mixed carbon sources.** Here we consider all possibilities for the choice of the biochemical parameters ($k_{cat}$ can be any positive values). Predictions based on parameters listed in Supplementary Table 1 are shown in blue text (see Supplementary Table 4 for details). Other patterns that qualitatively similar with experiments are shown in orange text.

| Mixed Substrates | Possible Pattern Types | Pools a1, a2, a3, a4 | Pool b | Pool c | Pool d |
|---|---|---|---|---|---|
| | | Ser, Gly, Cys, Trp, Phe, Tyr, His; precursors of RNA, DNA, Glycogen, Lipoglycans, Murein. | Ala, Val, Leu, Ile*; precursors of Lipids | Glu, Gln, Pro, Arg | Asp, Asn, Met, Thr, Lys, Ile* |
| | | Suppliers | Suppliers | Suppliers | Suppliers |
| A (Glucose/ Lactose/ Fructose/ Glycerol/ Maltose/ Galactose) + B1 (Pyruvate) | DX* No.1 | A | A | A | A |
| | DX No.2 | B1 | B1 | B1 | B1 |
| | CoU* No.1 | A | B1 | A (33%), B1 (67%) | A (55%), B1 (45%) |
| | CoU No.2 | A | B1 | B1 | B1 |
| | CoU No.3 | A (a1, a2, a3), B1 (a4) | B1 | B1 | B1 |
| | CoU No.4 | A (a1, a2), B1 (a3, a4) | B1 | B1 | B1 |
| | CoU No.5 | A(a1), B1 (a2, a3, a4) | B1 | B1 | B1 |
| | CoU No.6 | A | A (50%), B1 (50%) | A (66.5%), B1 (33.5%) | A (77.5%), B1 (22.5%) |
| | CoU No.7 | A | A (50%), B1 (50%) | A (50%), B1 (50%) | A (50%), B1 (50%) |
| | CoU No.8 | A | B1 | A (16.5%), B1 (83.5%) | A (27.5%), B1 (72.5%) |
| | CoU No.9 | A (a1, a2, a3, 50% a4), B1 (50% a4) | B1 | A (16.5%), B1 (83.5%) | A (27.5%), B1 (72.5%) |
| | CoU No.10 | A (a1, a2, a3, 50% a4), B1 (50% a4) | B1 | B1 | B1 |
| | CoU No.11 | A (a1, a2, 50% a3), B1 (a4, 50% a3) | B1 | B1 | B1 |
| | CoU No.12 | A (a1, 50% a2), B1 (a3, a4, 50% a2) | B1 | B1 | B1 |
| | CoU No.13 | A (50% a1), B1 (a2, a3, a4, 50% a1) | B1 | B1 | B1 |
| A (Glucose/ Lactose/ Fructose/ | DX No.1 | A | A | A | A |
| | DX No.2 | B2 | B2 | B2 | B2 |
| | CoU No.1 | A | A | A (67%), B2 (33%) | A (45%), B2 (55%) |
| Glycerol/<br>Maltose/<br>Galactose)<br>+B2<br>(Succinate/<br>Malate/<br>Fumarate) | CoU No.2 | A | B2 | B2 | B2 |
| | CoU No.3 | A | B2 | A (33%),<br>B2 (67%) | A (55%),<br>B2 (45%) |
| | CoU No.4 | A (a1, a2, a3), B2 (a4) | B2 | B2 | B2 |
| | CoU No.5 | A (a1, a2), B2 (a3, a4) | B2 | B2 | B2 |
| | CoU No.6 | A (a1), B2 (a2, a3, a4) | B2 | B2 | B2 |
| | CoU No.7 | A | A (50%),<br>B2 (50%) | A (66.5%),<br>B2 (33.5%) | A (77.5%),<br>B2 (22.5%) |
| | CoU No.8 | A | A | A (83.5%),<br>B2 (16.5%) | A (72.5%),<br>B2 (27.5%) |
| | CoU No.9 | A | A (50%),<br>B2 (50%) | A (33.5%),<br>B2 (66.5%) | A (22.5%),<br>B2 (77.5%) |
| | CoU No.10 | A | B2 | A (16.5%),<br>B2 (83.5%) | A (27.5%),<br>B2 (72.5%) |
| | CoU No.11 | A | A (50%),<br>B2 (50%) | A (50%),<br>B2 (50%) | A (50%),<br>B2 (50%) |
| | CoU No.12 | A (a1, a2, a3, 50% a4),<br>B2 (50% a4) | B2 | B2 | B2 |
| | CoU No.13 | A (a1, a2, a3, 50% a4),<br>B2 (50% a4) | A (50%),<br>B2 (50%) | A (33.5%),<br>B2 (66.5%) | A (22.5%),<br>B2 (77.5%) |
| | CoU No.14 | A (a1, a2, a3, 50% a4),<br>B2 (50% a4) | B2 | A (16.5%),<br>B2 (83.5%) | A (27.5%),<br>B2 (72.5%) |
| | CoU No.15 | A (a1, a2, 50% a3),<br>B2 (a4, 50% a3) | B2 | B2 | B2 |
| | CoU No.16 | A (a1, 50% a2),<br>B1 (a3, a4, 50% a2) | B2 | B2 | B2 |
| | CoU No.17 | A (a1, 50% a2),<br>B1 (a3, a4, 50% a2) | B2 | B2 | B2 |
| B1<br>(Pyruvate)<br>+B2<br>(Succinate/<br>Malate/<br>Fumarate) | DX No.1 | B1 | B1 | B1 | B1 |
| | DX No.2 | B2 | B2 | B2 | B2 |
| | CoU No.1 | B1 | B1 | B1 (67%),<br>B2 (33%) | B1 (45%),<br>B2 (55%) |
| | CoU No.2 | B2 | B1 | B1 (67%),<br>B2 (33%) | B1 (45%),<br>B2 (55%) |
| | CoU No.3 | B1 | B1 | B1 (83.5%),<br>B2 (16.5%) | B1 (72.5%),<br>B2 (27.5%) |
| | CoU No.4 | B2 | B1 (50%)<br>B2 (50%) | B1 (33.5%),<br>B2 (66.5%) | B1 (22.5%),<br>B2 (77.5%) |
| | CoU No.5 | B1 (50%), B2 (50%) | B1 | B1 (67%),<br>B2 (33%) | B1 (45%),<br>B2 (55%) |
|   | CoU No.6 | B1 (50%), B2 (50%) | B1 (50%), B2 (50%) | B1 (33.5%), B2 (66.5%) | B1 (22.5%), B2 (77.5%) |
|   | CoU No.7 | B1 (50%), B2 (50%) | B1 | B1 (83.5%), B2 (16.5%) | B1 (72.5%), B2 (27.5%) |
|   | CoU No.8 | B1 (50%), B2 (50%) | B1 (50%), B2 (50%) | B1 (50%), B2 (50%) | B1 (50%), B2 (50%) |
| Succinate + Malate | DX No.1 | Succinate | Succinate | Succinate | Succinate |
|   | DX No.2 | Malate | Malate | Malate | Malate |
|   | CoU No.1 | Succinate (50%), Malate (50%) | Succinate (50%), Malate (50%) | Succinate (50%), Malate (50%) | Succinate (50%), Malate (50%) |

[*]DX denotes diauxie and CoU signifies co-utilization.